\RequirePackage{fix-cm}
\documentclass[smallextended]{svjour3}       % onecolumn (second format)
\smartqed  % flush right qed marks, e.g. at end of proof
\usepackage{latexsym}
\usepackage{graphicx}
\usepackage{cuted}
\usepackage[a4paper,left=2.5cm,right=2.5cm,top=2.5cm,bottom=2.5cm]{geometry}
\usepackage{amsmath}
\usepackage{amsfonts}
\usepackage{amssymb}
\usepackage[utf8]{inputenc}
\usepackage[pagebackref]{hyperref}
\usepackage{epigraph} 
\journalname{}
\begin{document}

\title{A reconstruction of quantum theory for spinning particles}
%old title: Projecting classical probabilistic ensembles to spinning quantum ensembles


\author{U. Klein}


\institute{U. Klein \at
              University of Linz \\
              Institute for Theoretical Physics\\
              A-4040 Linz, Austria\\
              Tel.: 0043 660 5489735\\
              \email{ulf.klein@jku.at}           %  \\
}

\date{Received: date / Accepted: date}

\maketitle

\begin{abstract}
As part of a probabilistic reconstruction of quantum theory (QT), we show that spin is not a purely
quantum mechanical phenomenon, as has long been assumed. Rather, this phenomenon occurs
before the transition to QT takes place, namely in the area of the quasi-classical (here
better quasi-quantum) theory. This borderland between classical physics and QT can be reached
within the framework of our reconstruction by the replacement $p \rightarrow M (q, t)$,
where $p$  is the momentum variable of the particle and $M(q, t)$ is the momentum field in
configuration space. The occurrence of spin, and its special value $1/2$ , is a consequence of
the fact that $M(q,t)$ must have exactly three independent components $M_{k}(q,t)$ for a single
particle because of the three-dimensionality of space. In the Schr\"odinger equation for
a ``particle with spin zero'', the momentum field is usually represented as a  gradient of a
single function $S$. This implies dependencies between the components $M_{k}(q,t)$ for which
no explanation exists. In reality, $M(q,t)$ needs to be represented by three functions, two of
 which are rotational degrees of freedom. The latter are responsible for the existence of spin.
 All massive structureless particles in nature must therefore be spin-one-half particles, simply
 because they have to be described by $4$ real fields, one of which has the physical meaning
 of a probability density, while the other three are required to represent the momentum field in
 three-dimensional space. We derive the Pauli-Schr\"odinger equation, the correct value $g=2$
 of the gyromagnetic ratio, the classical limit of the Pauli-Schr\"odinger equation, and
 clarify some other open questions in the borderland between classical physics and QT.
\end{abstract}

\section{Introduction}
\label{intro}
\epigraph{"Niels Bohr brainwashed a whole generation of theorists into thinking that the job (interpreting quantum theory) was done 50 years ago''}{\textit{Murray Gell-Mann (1929-2019)}}Quantum theory (QT) is the
subject of controversial discussions since its invention more then ninety years ago.
The phenomenon of spin, discovered at about the same time, plays a particularly
mysterious role, because of the disturbing fact that we are unable to identify a
classical counterpart for this "classically non-describable two-valuedness"~\cite{pauli:coll-influence}.
One suspects that a satisfying explanation of spin will not be possible without a better
understanding of QT as a whole. There are two fundamentally different
ways to interpret the formalism of QT. The first way, mainly propagated by Bohr,
claims that QT is able to describe individual particles. In contrast to
this ``individuality interpretation'', Einstein's ``ensemble interpretation'' asserts that
QT is only able to describe statistical ensembles~\cite{ballentine:statistical}. It may seem
strange that a theory whose dynamical predictions are statistical in nature should be able
to describe individual particles. As a matter of fact, however, Bohr’s interpretation is
predominant at the moment; it is accompanied by ongoing discussions about various
internal contradictions.

The reference theory for any interpretation of QT must be part of classical physics. In
the context of the individuality interpretation, this reference theory is classical mechanics (CM).
Due to the fundamental differences in the mathematical formulation of both theories (ordinary
differential equations and observables defined as ordinary functions in CM, as opposed to partial
differential equations and observables defined as operators in QT), the similarities between QT
and CM are limited to structural (mathematical) similarities between both theories. A satisfactory
understanding of QT in this formal sense would be achieved if it were possible to map the
structural properties of classical observables in a one-to-one manner to corresponding properties
of quantum mechanical operators. However, such a general process called ``quantization''
does not exist, as Groenewold has shown~\cite{groenewold:principles}.

In the probabilistic version of CM, which we call probabilistic mechanics (PM), the object to
be examined is not one (or several) individual particles, but a statistical ensemble of (one or
several) particles. The observable quantities are expectation values, and the fundamental
dynamical variable, the probability density $\rho$, satisfies a partial differential equation (the Liouville
equation). Thus, in contrast to CM, the mathematical formulation of PM is very similar to that of QT.
This similarity allows for a completely different approach when studying the relation between QT
and classical physics, namely a ``derivation'' or ``reconstruction'' of one of the two theories from the
other. The relevance of the probabilistic component for a proper understanding of QT was
recognized very early by Van Vleck~\cite{vanvleck:correspondence},
Schiller~\cite{schiller:quasiclassical},\cite{schiller:spinning},
Bopp~\cite{bopp:mecanique},\cite{bopp:principles} and others.

This work is the fourth in a series of papers that attempt to understand QT on the basis of
Einstein’s ensemble interpretation. These works will be referred to
as I~\cite{klein:koopman}, II~\cite{klein:probabilistic}, and III~\cite{klein:nonspinning}.
According to the ensemble interpretation, the natural starting point for a reconstruction of
QT is PM. In I-III it was shown that it is possible, by choosing this starting point, to derive
not only Schr\"odinger's equation, but also most other fundamental properties of QT,
such as the non-commutative structure of operators, Born's rule and others. In the
present work it will be shown that even the strange phenomenon of spin may be
derived using  a refined version (actually a corrected version) of the theory
presented in III. 

The mathematical similarity between both theories allows a derivation of QT from PM  that
is very simple, at least in conceptual terms. The main formal difference between PM and QT
is the number of independent variables, as can be seen immediately if one descends from
Hilbert space to the most concrete formulation of QT in terms of differential equations.
Obviously, PM is a theory in phase space with, neglecting time at the moment, $2n$ independent
variables, while QT is a theory in configuration space with $n$ independent variables. Thus,
a reconstruction of QT must necessarily contain a projection from phase space to configuration
space. As a second fundamental step a linearization or randomization  must be performed,
as explained in detail in III. A theory containing these two steps, regardless of the order, is
referred to as ``Hamilton-Liouville-Lie-Kolmogorov theory'' (HLLK).

In works I and II, the first version of the HLLK was used, in which the linearization is performed
first followed by the projection. In III and the present work, the projection is performed first and
then the linearization (or randomization) follows as second step. This second version
of the HLLK allows for a deeper understanding of the relationship between QT and classical
physics. Let us also note, that in II a variety of different observables $A$ was investigated, while
in III and the present work we study only the particular sector $A=H$ of HLLK related to spin. 

The present work follows the same general scheme as III, but the actual implementation is
more involved. In the following section~\ref{sec:basic-equations} we introduce, as in III, a
general momentum field. As was shown in III, it is not possible to use the components $M_{k}$ 
of the momentum field themselves as dynamical variables because they are in general not
functionally independent from each other. Instead, one must take as dynamical variables
certain independent functions, called potentials, which may be used to represent the momenta;
the situation is similar as in QT. In III only irrotational momentum fields were dealt with,
which means that all components $M_{k}$ could  derived from a single function $S$. In this
case the transition from the $M_{k}$ to the new dynamic variable $S$ (which later becomes
the quantum mechanical phase) is obvious. In the present work we need a larger number
of potentials $S, Q_{\alpha}, P_{\alpha}$ in order to represent general momentum fields, that also
have vortical components. The fields $Q_{\alpha}, P_{\alpha}$ called Clebsch potentials describe
these vortical components and are ultimately responsible for the spin. The equations of motion for
the new dynamic variables $S, Q_{\alpha}, P_{\alpha}$ were derived by the mathematician
H. Rund~\cite{rund:clebsch}. In section~\ref{sec:evolution-equations} we give a brief overview
of Rund’s theory and then reproduce his results within the present formalism.

In the emerging quasi-quantal or quasi-classical theory, which is referred to as QA,
particle trajectories do still exist, but are only locally valid. In the following sections
we restrict ourselves to the case $N = 1$ of a single particle, where besides $S$
and $\rho$, only two additional (vortical) variables $Q,\,P$ occur. The QA (for $N=1$)
is structurally identical to the theory of inviscid barotropic fluids; this correspondence
is only of a formal nature and it should be born in mind that the physical meaning of
the variables is quite different. We use some results from fluid mechanics concerning
vortex lines. In section~\ref{sec:impl-topol-constr} we implement a topological restriction
for the variables $Q,\,P$. We assume that the mapping of $\mathbb{R}^{3}$  onto
the space of the $Q,\,P$ describes linked vortex lines (the linking number is a topological
invariant). The suitable mapping is given by the famous Hopf map, which implies the
replacement of the $Q,\,P$ with new \emph{canonical} Clebsch variables $\vartheta$
and $\varphi$. This map also provides the definition of a suitable two-component state
variable (spinor) $\psi$ which is of central importance for the transition to QT and for
the implementation of the minimal-coupling rule. As shown in
section~\ref{sec:intr-gauge-field} the special form of Hopf’s map is closely related
to the fact that the parameter space of the rotation group is doubly connected.

Using the new state variable $\psi$, we construct in section~\ref{sec:semil-evol-equat}
a semi-linear form of the evolution equations, which differs from the Pauli-Schrödinger
equation (without external electromagnetic field) only by a non-linear term. Only when
using this form of the evolution equations is it possible (see section~\ref{sec:intr-gauge-field})
to implement the minimal-coupling-rule in a natural way and to derive the correct g-factor of $2$ .
Section~\ref{sec:intr-gauge-field} also contains a discussion of the concept of the "magnetic
moment of the electron".

The transition to QT  is carried out in section~\ref{sec:tran-quant-theo-lin} by linearization with
respect to the variable $\psi$. This can be done very easily by eliminating the non-linear term
in the semi-linear equation. Section~\ref{sec:tran-quant-theo-lin} also examines which terms have
to be added to the original evolution equations for $\rho,\,S,\,\vartheta,\,\varphi$ in order to
generate the linearization.

The justification for performing the linearization, discussed in more detail in III, is that a large
number of globally valid solutions can only be generated in a linear theory (because of the
superposition principle). This point is discussed in more detail in III. While this reasoning is
understandable, the process of eliminating a term in a differential equation may seem 
somewhat formal, if not crude. One wonders what exactly happened during the discontinuous
transition from the quasi-classical equations to the quantum equations. In
section~\ref{sec:tran-quant-theo-rand} this discontinuous process is ``resolved'' and a
statistical theory is constructed which is as similar as possible to the quasi-classical theory,
but which does not claim to describe individual particles at all. This theory, whose basic
assumptions are the same as in III, represents the most detailed version of the HLLK.

In the last section~\ref{sec:discussion} we list the results obtained so far, taking into account both
versions of the HLLK, both the first version used in I and II and the second version used in III and
here. Then we go into more detail on some important points. In particular, we emphasize that
quantum spin cannot be a localizable property of individual particles, but a collective property of
the probabilistic single-particle ensemble, which is related to the possibility of rotations in
three-dimensional space. The three-dimensionality of space is also the reason for the special
value $1/2$  of the spin. Other points that will be discussed are the classical limit of the Pauli equation
and the role of potentials in quantum theory. We conclude with a few remarks concerning the
interpretation of the quantum theoretical formalism.
\section[Basic equations in phase space]{Basic equations}
\label{sec:basic-equations}
In this section we briefly recall all those relations from III that we need for our spin theory; detailed
explanations may be found in III.  The state of a system of $N $
particles is described by $n=3N$ coordinates $q={q_{1},...,q_{n}}$ and $n$ conjugate momenta
$p={p_{1},...,p_{n}}$. Particle trajectories $q_{k}(t),\,p_{k}(t)$ are given by the solutions of the
canonical equations
\begin{equation}
  \label{eq:SU2MDVI9ER}
\dot{q_{k}}=\frac{\partial H(q,p)}{\partial p_{k}},\;\;\;\dot{p}_{k}=-\frac{\partial H(q,p)}{\partial q_{k}}
\mbox{,}
\end{equation}
with Hamilton's function $H(q,p)$ not depending on time $t$.  A particular trajectory in the
$2n$-dimensional phase space $\Omega=\mathbb{R}^{n}_{q} \times \mathbb{R}^{n}_{p}$ may
be labeled by its state $q^{0},\,p^{0}$ at an initial time $t_{0}$. The solutions of~\eqref{eq:SU2MDVI9ER}
are written in the form
\begin{equation}
  \label{eq:GFRM2DSD08R}
q_{k}=Q_{k}(t,t^{0},q^{0},p^{0}) ,\;\;\;p=P_{k}(t,t^{0},q^{0},p^{0})
\mbox{,}
\end{equation}
where the dependence on $t_{0}$ will often be supressed. The variables $q,\,p$ are 
``Lagrangian coordinates'' representing particle properties. A statistical ensemble is
defined as the uncountable set of all solutions~\eqref{eq:GFRM2DSD08R}. Describing
ensembles with the help of ``Eulerian coordinates'', which are denoted by the same
symbols $q,\,p$ but represent points of $\Omega$, is much more convenient.  The most
important Eulerian dynamic variable is the probability density $\rho(q,p,t)$, ruling the
time-dependent distribution of trajectories in phase space, which obeys the Liouville equation
\begin{equation}
  \label{eq:ALI5OUVI2LE}
\frac{\partial \rho}{\partial t} + \frac{\partial \rho}{\partial q_{k}} \frac{\partial H}{\partial p_{k}}  
- \frac{\partial \rho}{\partial p_{k}} \frac{\partial H}{\partial q_{k}}=0
\mbox{.}
\end{equation}
In order to perform the transition from probabilistic mechanics (PM) to QT  we need a second Eulerian
variable, the action variable $S(q,p,t)$, which obeys the differential equation 
\begin{equation}
  \label{eq:DRAE48ZU9EB}
\frac{\partial S}{\partial t} + \frac{\partial S}{\partial q} \frac{\partial H}{\partial p}  
- \frac{\partial S}{\partial p} \frac{\partial H}{\partial q} = \bar{L}
\mbox{,}
\end{equation}
which will be referred to as action equation.  The quantity $\bar{L}$ (the Lagrangian) is defined by
\begin{equation}
  \label{eq:HG218I8LWE}
\bar{L}=\bar{L}(q,p,t)=p\,\frac{\partial H(q,p)}{\partial p} - H(q,p)
\mbox{.}
\end{equation}

As a first step on our way from PM to QT, we have to perform the projection of the basic
equations~\eqref{eq:SU2MDVI9ER},~\eqref{eq:ALI5OUVI2LE},~\eqref{eq:DRAE48ZU9EB}
onto configuration space.  We replace the particle momentum $p$ at each instant of time $t$
by a $n-$component momentum field $M$:   
\begin{equation}
  \label{eq:ADRAROGF}
p_{k} \rightarrow M_{k}(q,t)
\mbox{.}
\end{equation}
Thus, $2n-$dimensional phase space is projected to a $n$-dimensional subspace
$\mathbb{M}=\left\{(q,p)\in \Omega\,|\, p=M(q)\right\}$ which is parametrized by the
configuration space coordinates $q_{k}$~\cite{mukunda:phase-space}.

The projection of the canonical equations~\eqref{eq:SU2MDVI9ER} leads to the
differential equations~\cite{rund:clebsch,kozlov:general}
\begin{gather}
\dot{q}_{k}=v_{k}(q,t)
\mbox{,}
\label{eq:UMW29ISJG}\\
\frac{\partial M_{i}(q,t)}{\partial t}+
\left[\frac{\partial M_{i}(q,t)}{\partial q_{l}}- \frac{\partial M_{l}(q,t)}{\partial q_{i}}\right]
v_{l}\left( q,t \right) = -
\frac{\partial}{\partial q_{i}} h\left( q,t\right)
\mbox{,}
\label{eq:NJZIM2BSEQU}
\end{gather}
where the fields  $v(q,t)$, $h(q,t)$ are defined  in terms of $H(q,p),\, V(q,p)$ by   
\begin{gather}
  \label{eq:HDU33GHR7N}
v_{k}(q,t) = V_{k}\left(q,M(q,t)\right)
\mbox{,}\\
\label{eq:HD18IGZTEN}
h(q,t) = H\left(q,M(q,t)\right)
\mbox{,}\\
  \label{eq:HGW48IEVMEN}
V_{k}(q,p) = \frac{\partial H(q,p)}{\partial p_{k}}
\mbox{.}
\end{gather}  
Eq.~\eqref{eq:NJZIM2BSEQU} will be referred to as canonical condition. As a result of the projection
to configuration space the $2n$ ordinary differential equations~\eqref{eq:SU2MDVI9ER}
are replaced by $n$ ordinary differential equations~\eqref{eq:UMW29ISJG}, for the particle
positions $q$, and $n$  partial differential equations~\eqref{eq:NJZIM2BSEQU}, for the
momentum field $M$.

A useful quantity, characterizing  the purely rotational part of the momentum field, is the
vorticity tensor $\Omega_{ik}$, defined by 
\begin{equation}
  \label{eq:VO2OIT9TWE}
\Omega_{ik}(q,t):=\frac{\partial M_{k}(q,t)}{\partial q_{i}}-\frac{\partial M_{i}(q,t)}{\partial q_{k}}
  \mbox{.}
\end{equation}
Its  equation of motion, which may easily be derived from the canonical condition, shows that
the time-dependence of $\Omega_{ik}$ is completely determined by the solutions of the particle 
equations of motion~\eqref{eq:UMW29ISJG}.  As a consequence, performing a standard calculation
from fluid mechanics, one  may show that 
\begin{equation}
  \label{eq:GRI25OP9WF}
\Omega_{ik}(q,t)=\Omega_{lj}^{0}(q)\,\frac{\partial q_{l}^{0}(q,t)}{\partial q_{i}}
\,\frac{\partial q_{j}^{0}(q,t)}{\partial q_{k}}
\mbox{.}
\end{equation}
Here, the functions $q_{l}^{0}(q,t)$ are obtained by inverting the solutions $q_{l}(t,q^{0})$ 
of~\eqref{eq:UMW29ISJG}; the quantities $q^{0}$ and $\Omega_{lj}^{0}(q)$ are the initial values 
of $q_{l}(t)$ and $\Omega_{ik}(q,t)$. This relation shows explicitly that the time dependence 
of the vorticity tensor is determined by the flow of the particle equations of motion. In particular,
$\Omega_{lj}^{0}(q)=0$ implies $\Omega_{ik}(q,t)=0$ for all future times.

The projection of the Liouville equation to configuration space leads to the continuity relation 
\begin{equation}
  \label{eq:HSE27WG6FDC}
\frac{\partial \rho(q,t)}{\partial t} + \frac{\partial }{\partial q_{k}}\rho(q,t)
v_{k}(q,t)=0
\mbox{}
\end{equation}
for the probability density in configuration-space $\rho(q,t)$. The latter is defined by the relation
$\rho(q,p,t)=\rho(q,t) \delta(p-M(q,t))$, where $\delta$ is the $n$-dimensional delta function. The
projection of the action equation to configuration space may be put in the form
\begin{equation}
  \label{eq:NN8M43UQU}
\frac{\partial s}{\partial t}
+v_{k}\left[\frac{\partial s}{\partial q_{k}}-M_{k}(q,t) \right]+h=0
\mbox{, }
\end{equation}  
where the action field on configuration space $s(q,t)$ is defined by
\begin{equation}
  \label{eq:HID5ESC9SF}
s(q,t)=S(q,M(q,t),t)
  \mbox{.}
  \end{equation}
Relation~\eqref{eq:NN8M43UQU}  may also be obtained by means of a Lagrangian to
Eulerian transition of the action integral in configuration space.

As shown in III, the $n$ components $M_{1}(q,t),..,M_{n}(q,t)$  of the momentum field are generally
not functionally independent from each  other. They are therefore unsuitable as dynamic variables
and must be replaced by suitable, functionally independent quantities.
The problem of finding these new variables for arbitrary vector fields was solved by
Pfaff~\cite{caratheodory:calculus_I}.  For the present nonrelativistic problem, we need the part
of Pfaff’s solution relating to fields with an odd number $L$ of independent functions. In this case
there is an integer $m$ given by $L=2m+1$ and $M$ may be written in the form
\begin{equation}
  \label{eq:REP23HZU3DF}
M_{k}(q,t)=\frac{\partial S(q,t)}{\partial q_{k}}+
P_{\alpha}(q,t)\frac{\partial Q_{\alpha}(q,t)}{\partial q_{k}}
\mbox{,}
\end{equation}
where the Clebsch potentials~\cite{clebsch:transformation} $S(q,t),\,P_{\alpha}(q,t),\,Q_{\alpha}(q,t)$
are $2m+1$ independent functions of $q_{1},..,q_{n},t$ (Greek indices $\alpha,\,\beta,..$ run from $1$
to $m$ and double occurrence of these indices entails a summation from $1$ to $m$).
Inserting~\eqref{eq:REP23HZU3DF} in the definition~\eqref{eq:VO2OIT9TWE} we see that
the components 
\begin{equation}
  \label{eq:VOR3ZOI8NS}
\Omega_{ij}=\frac{\partial P_{\alpha}}{\partial q_{j}}\frac{\partial Q_{\alpha}}{\partial q_{i}}-
\frac{\partial Q_{\alpha}}{\partial q_{j}}\frac{\partial P_{\alpha}}{\partial q_{i}}
\mbox{.}
\end{equation}
of the vorticity tensor will generally be different from zero. As a consequence, if we consider a closed loop
$\gamma$, lying in $\mathbb{M}$ at time $t$, then the closed path integral
\begin{equation}
  \label{eq:A71ZU8WP2O}
\oint_{\gamma}\mathrm{d}q_{k} M_{k}(q,t)
\mbox{,}
\end{equation}
will be different from zero. Thus, the components $P_{\alpha}(q,t),\,Q_{\alpha}(q,t)$ of a
general momentum field ($m>0$) describe a \emph{vortical} state of motion of the
probabilistic ensemble; they will sometimes be referred to as ``vortical variables''.
In contrast, for the irrotational momentum fields considered in III the
integral~\eqref{eq:A71ZU8WP2O} vanishes for arbitrary $t$ (assuming that $S(q,t)$ is 
single-valued).

In III the case $m=0$ of an irrotional momentum field was studied. This case is unphysical but
important because it leads to Schr\"odinger's equation. The determination of the equation of
motion for the single Clebsch potential $S$ in III was simple because both the canonical condition
and the action equation reduce in this case to a Hamilton-Jacobi equation. In the following, some basic
results for the general case $m> 0$ are reported first; the transition to QT is carried out later
for $m = 1$. 

\section[Evolution equations]{Evolution equations}
\label{sec:evolution-equations}
Our task is to derive equations of motion for the new variables $q_{k}$,  $S(q,t)$,  $P_{\alpha}(q,t)$,
$Q_{\alpha}(q,t)$ from the equations of motion~\eqref{eq:UMW29ISJG},~\eqref{eq:NJZIM2BSEQU}
for the old variables $q_{k}$, $M_{k}(q,t)$. This is an easy task as regards the particle equation of
motion~\eqref{eq:UMW29ISJG}; one only has to insert the expansion~\eqref{eq:REP23HZU3DF} in
Eq.~\eqref{eq:UMW29ISJG},
\begin{equation}
  \label{eq:HRUM4R9N}
\dot{q}_{k}= V_{k}\left(q,\frac{\partial S(q,t)}{\partial q}+P(q,t)\frac{\partial Q(q,t)}{\partial q}\right)
\mbox{.}
\end{equation}
The less trivial problem of determining the evolution equations for the Clebsch potentials was
solved by the mathematician H. Rund more than four decades ago~\cite{rund:clebsch}.
Rund's original work contained an unnecessary restriction which was removed shortly afterwards
by Baumeister~\cite{baumeister:generalized}. This extension will of course be taken into account
in the present work. In this section we first give a brief overview and discussion of Rund’s theory
and then clarify the relationship between his variables and those used in this work.

\subsection[Outline of Rund's theory]{Outline of Rund's theory}
\label{sec:outline-runds-theory}
We give a brief outline of Rund’s theory for the convenience of the reader. 
If the expansion~\eqref{eq:REP23HZU3DF} is inserted in the canonical
condition~\eqref{eq:NJZIM2BSEQU} the resulting relation may be put in the form
\begin{equation}
  \label{eq:DD34NSA1DB}
\frac{\partial}{\partial q_{i}}T(q,t)+
\frac{\partial Q_{\alpha}}{\partial q_{i}} \frac{\mathrm{D}P_{\alpha}}{\mathrm{D}t}-
\frac{\partial P_{\alpha}}{\partial q_{i}} \frac{\mathrm{D}Q_{\alpha}}{\mathrm{D}t}=0
\mbox{,}
\end{equation}
where 
\begin{equation}
  \label{eq:SI34SA7TIB}
T(q,t) := H\left(q,\frac{\partial S}{\partial q}+ P \frac{\partial Q}{\partial q},t \right)+\frac{\partial S}{\partial t}+ P_{\alpha} \frac{\partial Q_{\alpha}}{\partial t}
\mbox{,}
\end{equation}
and the total derivatives of $P_{\alpha}$ and $Q_{\alpha}$ with respect to time  are defined by
\begin{equation}
\label{eq:FIRUZT2TSCHL}
\frac{\mathrm{D}}{\mathrm{D}t} =  
\frac{\partial}{\partial t}+
V_{l}\left(q,\frac{\partial S}{\partial q}+P \frac{\partial Q}{\partial q},t \right)
\frac{\partial}{\partial q_{l}}
\mbox{.}
\end{equation}
In fluid mechanics these total derivatives are referred to as convective derivatives or material
derivatives.

In the next step, one takes advantage of the functional independence of $S$, $P_{\alpha}$, 
$Q_{\alpha}$. As a consequence a one-to-one correspondence between the $n$ variables 
$S,\,P_{\alpha},\,Q_{\alpha},q_{2m+2},..,q_{n}$ and the $n$ variables $q_{1},..,q_{n}$ must exist 
at each instant of time. As a further consequence, a function $\Phi$, depending on 
$S,\,P_{\alpha},\,Q_{\alpha},q_{2m+2},..,q_{n}$ must exist which fulfills the relation
\begin{equation}
  \label{eq:DG4RVI9EGK}
\Phi(t,\,S,\,P_{\alpha},\,Q_{\alpha},q_{2m+2},..,q_{n})=T(q_{1},..,q_{n},t)
\mbox{.}
\end{equation}
In two further important steps it is shown that $\Phi$ depends neither on $S$ nor on $q_{2m+2},..,q_{n}$; 
the latter result is required for $2m+1 < n$~\cite{baumeister:generalized}. The final result of  
Rund's theory is given by the dynamic equations 
\begin{gather}
\frac{\partial S}{\partial t}+ 
P_{\alpha} \frac{\partial Q_{\alpha}}{\partial t}+
H\left(q,\frac{\partial S}{\partial q} + P \frac{\partial Q}{\partial q},t \right)-
\Phi(Q,\,P,\,t)  =  0
\mbox{,}
\label{eq:KUZ2RETJZB}\\
\frac{\mathrm{D}P_{\alpha}}{\mathrm{D}t}  =  
-\frac{\partial \Phi(Q,\,P,\,t)}{\partial Q_{\alpha}}
\mbox{,}
\label{eq:FIR2JKKRCHL}\\
\frac{\mathrm{D}Q_{\alpha}}{\mathrm{D}t} = 
\frac{\partial \Phi(Q,\,P,\,t)}{\partial P_{\alpha}} 
\label{eq:SJZTE3SCHL}
\mbox{,}
\end{gather}
where the total derivatives are defined by~\eqref{eq:FIRUZT2TSCHL}. The function 
$\Phi(Q,\,P,\,t)$ is completely arbitrary.

Besides these fundamental equations we quote from Rund's work the following important
relation which is obtained by differentiating Eq.~\eqref{eq:KUZ2RETJZB} with respect to $q_{k}$:
\begin{equation}
  \label{eq:AWEOI34T7RM}
\left[\frac{\mathrm{D}M_{k}}{\mathrm{D}t}+\frac{\partial H}{\partial q_{k}}\right]
-\frac{\partial Q_{\alpha}}{\partial q_{k}}
\left[\frac{\mathrm{D}P_{\alpha}}{\mathrm{D}t}+\frac{\partial \Phi}{\partial Q_{\alpha}} \right]
+\frac{\partial P_{\alpha}}{\partial q_{k}}
\left[\frac{\mathrm{D}Q_{\alpha}}{\mathrm{D}t}-\frac{\partial \Phi}{\partial P_{\alpha}} \right]
=0
\mbox{.}
\end{equation}
Equation~\eqref{eq:KUZ2RETJZB} is referred to by Rund as "generalized Hamilton-Jacobi equation". 
Indeed, if we set $Q_{\alpha}=P_{\alpha}=\Phi=0$, Eq.~\eqref{eq:KUZ2RETJZB} reduces to the
Hamilton-Jacobi equation. Equations~\eqref{eq:FIR2JKKRCHL},~\eqref{eq:SJZTE3SCHL} are 
referred to as "associated canonical equations", as they may be interpreted as ordinary differential 
equations of canonical form, with a ``Hamiltonian'' $\Phi(Q,\,P,\,t)$, for the ``particle variables'' 
$P_{\alpha}(t)=P_{\alpha}(q(t),t),\,Q_{\alpha}(t)=Q_{\alpha}(q(t),t)$.\\[0.3cm]
As a first step towards a physical interpretation of these equations we note the following points:
\begin{itemize}
\item The ``Hamiltonian'' $\Phi$ is arbitrary; no physical result can depend on
its functional form. We are therefore allowed to set $\Phi=0$. Rund has already
shown, using his theory of Clebsch gauge transformations, that a gauge with vanishing
$\Phi$ may be introduced~\cite{rund:clebsch}; see also~\cite{sudarshan:classical}.
\item The relations~\eqref{eq:FIR2JKKRCHL},~\eqref{eq:SJZTE3SCHL} reduce for $\Phi=0$
  to the condition that the potentials $Q_{\alpha},\,P_{\alpha}$ are constant if transported
  along the velocity field
$V\left(q,\frac{\partial S}{\partial q}+P\frac{\partial Q}{\partial q},t\right)$. 
\item Equation~\eqref{eq:AWEOI34T7RM} shows that the generalized Hamilton-Jacobi 
equation~\eqref{eq:KUZ2RETJZB} makes sure that the validity 
of~\eqref{eq:FIR2JKKRCHL},~\eqref{eq:SJZTE3SCHL} implies the validity of the second canonical 
equation [see Eqs.~\eqref{eq:SU2MDVI9ER}] and vice versa.  
\end{itemize}
In view of the important role of the Hamilton-Jacobi equation for the quantum-classical
transition of the Schr\"odinger equation, the physical meaning of the generalized
Hamilton-Jacobi equation, in particular the meaning of the new variables $Q_{\alpha},\,P_{\alpha}$,
is of great interest. Its possible role with regard to QT has, however, not been clarified,
neither by Rund himself nor by authors elaborating later on his
theory~\cite{baumeister:generalized,samuel:phase_space}. Samuel, who constructed an elegant
phase-space version of Rund's theory wrote~\cite{samuel:phase_space}: 
\begin{quote}
It is not clear at the moment whether the generalized theory has any relevance to quantum
mechanics. It does seem safe to say however that analogies with quantum mechanics, if they exist,
are not straightforward and will require some unearthing.      
\end{quote}
In section~\ref{sec:tran-quant-theor} it will be shown that Rund's theory, as completed by the
continuity equation, leads for $N=1$ to the quasi-classical counterpart of the quantum theory of a
single particle (ensemble) with spin.

\subsection[Implementing the Clebsch potentials in the ensemble equations]{Implementing
  the Clebsch potentials in the ensemble equations}
\label{sec:intr-clebsch-ensemb}
We now use Pfaff’s expansion~\eqref{eq:REP23HZU3DF} to rewrite the ensemble equations in
configuration space~\eqref{eq:HSE27WG6FDC},~\eqref{eq:NN8M43UQU}  in terms of the new variables
$S,\,Q_{\alpha},\,P_{\alpha}$. We expect to reproduce  Rund’s  results and also to find relations
between the quantities $S(q,p,t)$, $S(q,t)$, and $s(q,t)$.

The continuity equation~\eqref{eq:HSE27WG6FDC} can be simply rewritten by
inserting the expansion~\eqref{eq:REP23HZU3DF}, 
\begin{equation}
  \label{eq:CHOTR7WW3FDC}
\frac{\partial \rho(q,t)}{\partial t} + \frac{\partial }{\partial q_{k}}\rho(q,t)
V_{k}\left(q,\frac{\partial S}{\partial q}+P\frac{\partial Q}{\partial q},t\right)=0
\mbox{.}
\end{equation}
The continuity equation~\eqref{eq:CHOTR7WW3FDC}, the generalized Hamilton Jacobi
equation~\eqref{eq:KUZ2RETJZB}, and the associated canonical equations~\eqref{eq:FIR2JKKRCHL},
~\eqref{eq:SJZTE3SCHL} represent a closed system of $2m+2$ equations for the $2m+2$ variables
$\rho,\,S,\,Q_{\alpha},\,P_{\alpha}$. This set of extended Rund's equations (extended by the continuity
equation) is a precursor of the quasi-quantal approximation (QA) of PM (in the actual QA  the
variables $Q,\,P$ will be replaced by other variables due to a topological restriction; see
section~\ref{sec:impl-topol-constr}).   This is of course still a classical system of equations; this
means that the equations for $S,\,Q_{\alpha},\,P_{\alpha}$ form a closed system that does not
contain the variable $\rho$. Given a solution of these ``deterministic'' equations the motion
of particles may be determined with the help of~\eqref{eq:HRUM4R9N}. 
 
The projected action equation is given by Eq.~\eqref{eq:NN8M43UQU}, where $s(q,t)$ is now defined by    
\begin{equation}
  \label{eq:HTE3RAJS7SF}
s(q,t)=S\left(q,\frac{\partial S}{\partial q}+P\frac{\partial Q}{\partial q},t\right)
  \mbox{,}
  \end{equation}
in terms  of the phase space action $S(q,p,t)$. In order to rewrite~\eqref{eq:NN8M43UQU} we use Pfaff's
expansion~\eqref{eq:REP23HZU3DF} and introduce the symbol $\chi(q,t)=s(q,t)-S(q,t)$ for the difference
betweeen $s$ and $S$. The projected action equation  may then be written in the form 
\begin{equation}
\label{eq:KU6PARUM9B}
\frac{\mathrm{D}\chi}{\mathrm{D}t}
-P_{\alpha}\frac{\mathrm{D}Q_{\alpha}}{\mathrm{D}t}+
\frac{\partial S}{\partial t}+ 
P_{\alpha} \frac{\partial Q_{\alpha}}{\partial t}+
H\left(q,\frac{\partial S}{\partial q} + P \frac{\partial Q}{\partial q},t \right) =  0
\mbox{,}
\end{equation}
We calculate the derivative of~\eqref{eq:NN8M43UQU} with respect to $q_{i}$ and change the order 
of derivatives to obtain
\begin{equation}
  \label{eq:LZ7UI432QU}
\frac{\partial}{\partial t}\frac{\partial s}{\partial q_{i}}
+\frac{\partial}{\partial q_{i}}
v_{k}\left[\frac{\partial s}{\partial q_{k}}-M_{k}(q,t) \right]+
\frac{\partial}{\partial q_{i}} h=0
\mbox{.}
\end{equation}  
The canonical condition~\eqref{eq:NJZIM2BSEQU} and the action equation in the
form~\eqref{eq:LZ7UI432QU} describe basically the same physics. However,
Eq.~\eqref{eq:NJZIM2BSEQU} is an initial value problem for $M_{k}(q,t)$ while
Eq.~\eqref{eq:LZ7UI432QU} describes the relation between $M_{k}(q,t)$ and $s(q,t)$.
Both equations should agree if an appropriate representation of $M_{k}(q,t)$ is
chosen. This requirement leads to the following condition for the Clebsch potentials:
\begin{equation}
  \label{eq:AKJU342RE9M}
\frac{\partial}{\partial q_{i}}
\left[\frac{\mathrm{D}\chi}{\mathrm{D}t}-P_{\alpha}\frac{\mathrm{D}Q_{\alpha}}{\mathrm{D} t}\right]
-\frac{\partial Q_{\alpha}}{\partial q_{i}}\frac{\mathrm{D}P_{\alpha}}{\mathrm{D}t}
+\frac{\partial P_{\alpha}}{\partial q_{i}}\frac{\mathrm{D}Q_{\alpha}}{\mathrm{D}t}
=0
\mbox{.}
\end{equation}
This condition is fulfilled  if $\chi,\,Q_{\alpha},\,P_{\alpha}$ are solutions of  the associated
canonical equations Eqs.~\eqref{eq:FIR2JKKRCHL},~\eqref{eq:SJZTE3SCHL} and
$\chi$  fulfills the relation
\begin{equation}
  \label{eq:BG4RTP8OZU}
\frac{\mathrm{D}\chi}{\mathrm{D}t}-
P_{\alpha}
\frac{\partial \Phi}{\partial P_{\alpha}}+\Phi=0
\mbox{,}
\end{equation}
where $\Phi$ is an arbitrary function of $Q,\,P,\,t$. The generalized Hamilton-Jacobi equation
follows from~\eqref{eq:KU6PARUM9B} if Eq.~\eqref{eq:BG4RTP8OZU} is taken into account. 
The results of Rund’s theory are thus reproduced if  the ensemble equations are combined with
Pfaff’s expansion. 

Equation~\eqref{eq:BG4RTP8OZU} clarifies the relation between the Clebsch potential $S(q,t)$ and
the projected action $s(q,t)$. The difference $\chi=s-S$ plays the role of an action defined for a
``Hamiltonian'' $\Phi$ in a $2m$-dimensional phase space with coordinates $Q_{\alpha},\,P_{\alpha}$. 
We may set $\Phi=0$, thereby destroying the canonical structure of 
Eqs.~\eqref{eq:FIR2JKKRCHL},~\eqref{eq:SJZTE3SCHL},~\eqref{eq:BG4RTP8OZU}.
Condition~\eqref{eq:BG4RTP8OZU} reduces for $\Phi=0$ formally to the corresponding condition
in III for the irrotational case; note however that the velocity field is defined differently. 
Then, for this most important gauge, $\chi,\,Q_{\alpha},\,P_{\alpha}$ become quantities
``moving with the flow'', i.e. $\frac{\mathrm{D}\chi}{\mathrm{D}t}=0$,
$\frac{\mathrm{D}Q_{\alpha}}{\mathrm{D}t} =0$, $\frac{\mathrm{D}P_{\alpha}}{\mathrm{D}t} =0$.
The simplest solution of $\frac{\mathrm{D}\chi}{\mathrm{D}t}=0$ is again $\chi=0$, just as in
the irrotational case.

\subsection[Interpretation of Clebsch potentials]{Interpretation of 
  Clebsch potentials}
\label{sec:interpr-clebsch-pote}
Let us ask if the new dynamical degrees of freedom $S(q,t),\,P_{\alpha}(q,t),\,Q_{\alpha}(q,t)$ may
be understood in terms of any standard concept of physics. This question is not answered in
Rund’s theory. With regard to the Clebsch variable $S$, which is also responsible for the
$U (1)$ gauge mechanism, we can answer this question in the affirmative. We  derived in
section~\ref{sec:intr-clebsch-ensemb} the relation between $S (q, t)$ and the projected
action variable $s (q, t)$. The variable $S (q, t)$ can therefore be traced back to a standard
concept in phase space, namely the action variable $S (q, p, t)$. We mention without
proof that this connection can also be established within the framework of the theory
of canonical transformations; if we are given a complete solution of the Hamilton-Jacobi
equation, we can use it  to construct a projection onto the configuration space with prescribed
initial values. 

The question arises if the vortical variables $P_{\alpha}(q,t),\,Q_{\alpha}(q,t)$, may be understood
in an analogous way. Can we find a structure in phase space,  using a possibly extended theory of
canonical transformations, which allows us to understand these variables in a similar way as $S(q,t)$ ?
A closer look at the theory of canonical transformations shows that such an extension does
probably not exist. This implies that the variables $Q,\,P$ describe probably a structure which
has its origin in configuration space. A different concept, valid only in configuration space,
must be found if we want to understand the physical meaning of the variables $Q,\,P$. 

The following two hints, as to the physical origin of $Q,\,P$, may be obtained by looking more
closely at the basic equations~\eqref{eq:KUZ2RETJZB}-\eqref{eq:AWEOI34T7RM}
of Rund's theory:
\begin{itemize}
\item There are two different terms (both sums over $\alpha$) in Eq.~\eqref{eq:KUZ2RETJZB}
containing the Clebsch potentials. These terms appear exactly at the positions where the
electrodynamic scalar and vector potentials are located according to the minimal coupling rule.
\item  Relation~\eqref{eq:AWEOI34T7RM} shows that the appearance of the new variables does
not lead to new forces in the particle equations of motion. Introducing $Q,\,P$ just means
introducing new degrees of freedom, which do not interact with the particle coordinates. 
\end{itemize}
The first of these hints tells us that the vortical terms due to $Q,\,P$ may possibly be
interpreted as electrodynamic potentials, giving rise to nonvanishing electrodynamic fields.
The second hint tells us that these fields must be constructed in such a way that they do
\emph{not} exert any forces on the particles.  In a future publication it will be shown that
such ``internal potentials'' actually exist, and that the vortical Clebsch potentials may be
understood with the help of this concept.

\section[Implementing a topological constraint]{Implementing a topological constraint}
\label{sec:impl-topol-constr}
In the remaining part of this paper we will restrict ourselves to the most important case $N=1, \,n=3$
of a single particle ensemble. The number $m$ pairs of functions $P_{\alpha}(q,t),\,Q_{\alpha}(q,t)$
is equal to $1$ and the particle momentum is  specified at each space-time point $q,t$ by three
numbers $S(q,t),\,Q(q,t),\,P(q,t)$, corresponding to the fact that the momentum  must have three
independent components at each point of three-dimensional space. Further, the Clebsch gauge will be fixed
according to $\Phi=0$ and the Hamiltonian~\eqref{eq:IE78FA12HTA} without external electromagnetic
fields will be used (this important point will be taken into account in section~NN). The basic equations
for the four fields $S(q,t),\,Q(q,t),\,P(q,t),\,\rho(q,t)$ are then given by
\begin{gather}
\frac{\partial S}{\partial t}(q,t)+ 
P(q,t) \frac{\partial Q}{\partial t}(q,t)+
H^{0}\left(q,\frac{\partial S}{\partial q} + P \frac{\partial Q}{\partial q}\right)=  0
\mbox{,}
  \label{eq:LBRCS2CEW}\\
  \left[\frac{\partial}{\partial t}+V_{k}^{0}\left(q,\frac{\partial S}{\partial q} + P \frac{\partial Q}{\partial q}\right)\frac{\partial}{\partial q_{k}} \right] Q(q,t)=0
\mbox{,}
\label{eq:E7EGGALA9Q}\\
\left[\frac{\partial}{\partial t}+ V_{k}^{0}\left(q,\frac{\partial S}{\partial q} + P \frac{\partial Q}{\partial q}\right)\frac{\partial}{\partial q_{k}} \right] P(q,t)=0
\mbox{,}
\label{eq:T89ADU7I9Q}\\
\frac{\partial \rho}{\partial t}(q,t)+
\frac{\partial}{\partial q_{k}}\rho(q,t)V_{k}^{0}
\left(q,\frac{\partial S}{\partial q} + P \frac{\partial Q}{\partial q}\right)=0
\label{eq:T39AH4Q3UZ}
\mbox{.}
\end{gather}
The particle equations of motion $\dot{q}_{k}=v_{k}(q,t)$, with the velocity field defined 
by~\eqref{eq:HRUM4R9N}, are still valid in their limited range of validity. In this section we
consider a topological property of the mapping from $\mathbb{R}^{3}$ into the space of the $S, Q, P$,
which implies a more specific form of the $Q, P$. 

For the simplest case of a single free particle considered now, the relationship between momentum,
Clebsch potentials, and velocity is given by
\begin{equation}
  \label{eq:GRE23ZU7D4F}
M_{k}(q,t)=\partial_{k} S(q,t)+P(q,t)\partial_{k} Q(q,t)=mv_{k}(q,t)
\mbox{.}
\end{equation}
The vorticity tensor $\Omega_{ik}$ [see Eq.~\eqref{eq:VOR3ZOI8NS}] may conveniently be replaced
by an axial vorticity vector $\Omega_{i}=\frac{1}{2}\epsilon_{ikl}\Omega_{kl}=\epsilon_{ikl}\partial_{k}M_{l}$,
which is closely related to the vorticity $\omega_{i}=\epsilon_{ikl}\partial_{k}v_{l}$ defined by the velocity
field $v$.  The relationship between these vorticities and the Clebsch potentials is given by
 \begin{equation}
  \label{eq:HAPR23KR4D2F}
\Omega_{i}(q,t)= \epsilon_{ikl} \partial_{k}P(q,t)\partial_{l} Q(q,t)=m\omega_{i}(q,t)
\mbox{,}
\end{equation}
and the equation of motion for the three-vector $\mathbf{\Omega}$ may be written in the form 
\begin{equation}
  \label{eq:KR4UWA9MB}
\frac{\partial \mathbf{\Omega}}{\partial t}+\mathbf{\nabla} \times \left(\mathbf{\Omega} \times \mathbf{v}\right)=0
  \mbox{,}
\end{equation}
which is familiar from fluid mechanics. As mentioned already, there is a strong formal analogy between 
this part of the present theory and the theory of inviscid (ideal) barotropic fluids. This allows us to use
several important results from fluid mechanics in the present context; bearing always in mind the completely
different physical meaning of the variables. 

The vorticity field $\mathbf{\Omega}$ is by definition solenoidal. Surface effects do not exist in our theory as
we are considering an infinitely extended medium with sufficiently rapidly decreasing variables. The field
lines of $\mathbf{\Omega}$, referred to as vortex lines, are therefore closed curves. The topology of
a vector field is basically determined by the mutual position of its field lines, and in particular by the
interlinking of its field lines. This structure remains invariant under the time evolution given by
Eq.~\eqref{eq:KR4UWA9MB}. Helmholtz's and Kelvin's theorems, which also apply in the present theory,
can be interpreted in terms of the invariance of the topological structure. 

The case of linked vortex lines may be explained in terms of simple physical concepts
[see~\cite{moffatt:degree},~\cite{ranada:helicity} and references therein]. For example, in fluid
mechanics Moffatt considers two idealized vortex filaments $C_{1}$ and $C_{2}$ with
vanishing vorticity outside the filaments~\cite{moffatt:degree}. Stokes’s theorem leads then
immediately to the conclusion that the closed path integral along filament $C_{1}$ differs only from
zero if filament $C_{2}$ penetrates the surface spanned by $C_{1}$. Generalizing this consideration to
more realistic distributions he arrived at the conclusion that the quantity 
\begin{equation}
  \label{eq:DE4GHL9MMA}
H_{\omega}=\int  \mathrm{d}^{3}q \,\mathbf{v}\,\mathbf{\omega}
  \mbox{,}
  \end{equation}
referred to as (total) helicity, is a temporal and topological invariant, characterizing the degree of linkage
of vortex lines.

Using the present notation it is more convenient to work with  the helicity $H_{\Omega}=m^{2}H_{\omega}$, which
is defined by~\eqref{eq:DE4GHL9MMA} with  $v,\,\omega$  replaced by $M,\,\Omega$.
In order to study the variation of $H_{\Omega}$ with time, we need the evolution equation for the helicity density
$h_{\Omega}=M_{k}\Omega_{k}$. It is obtained by multiplying the canonical equation~\eqref{eq:NJZIM2BSEQU}
and the vorticity equation~\eqref{eq:KR4UWA9MB} by $\Omega$ and $M$ respectively, and adding both equations.
The result may be put in the form 
\begin{equation}
  \label{eq:DI1T7REVE8Q}
  \frac{\partial}{\partial t}h_{\Omega}+\mathbf{\nabla}
  \left(h \mathbf{\Omega} + h_{\Omega}\mathbf{V}-(\mathbf{M}\mathbf{V})\mathbf{\Omega}\right)
  =0
  \mbox{}
  \end{equation}
which shows explicitly the invariance of $H_{\Omega}$. The total helicity vanishes if linked vortex lines do
nowhere exist. On the other hand, if linked vortex lines exist $M$ cannot be single valued; in this case we
expect topological singularities of some kind. In order to examine this point more closely we calculate
$H_{\Omega}$ using Pfaff's formula~\eqref{eq:REP23HZU3DF}, expressing $M$ in terms of the Clebsch
potentials $S,\,P,\,Q$. Neglecting a surface integral at infinity the helicity may be written as
\begin{equation}
  \label{eq:NAS13ITH7MA}
  H_{\Omega}=-\int  \mathrm{d}^{3}q \,S\epsilon_{ikl}
  \left(\frac{\partial^{2}P}{\partial q_{i}\partial q_{k}}\frac{\partial Q}{\partial q_{l}}+\frac{\partial^{2}Q}{\partial q_{i}\partial q_{l}}\frac{\partial P}{\partial q_{k}} \right)
  \mbox{.}
  \end{equation}
 This formula shows that $H_{\Omega}$ vanishes if  $P,\,Q$ are both $C^{2}$.
 Linked vortex lines are therefore, as expected, related to singular, typically multivalued, behavior
 of the $P,\,Q$. In order to obtain a  nonvanishing $H_{\Omega}$ it is not required that the first order
 derivatives of both Clebsch potentials do not commute;  it suffices if only one of the  $P,\,Q$ is singular.

\subsection[The Hopf map]{The Hopf map}
\label{sec:hopf-map}
 The mapping of $\mathbb{R}^{3}$ in the space of the Clebsch variables $P,\,Q$, which is suitable for
 describing the quantum mechanical ensembles occurring in nature, belongs to a topologically non-trivial
 class, with linked vortex lines and non-vanishing helicity. It is given by the so-called Hopf  map or
 Hopf bundle~\cite{hopf:abbildungen}. We use the complex form of the Hopf map (see
 e.g.~\cite{socolovsky:spin}), which also provides us with an  appropriate definition for the new
 (spinorial) state variable, which will later be used to perform the transition to QT. Let $\mathbf{z}$
 denote the elements of the two-dimensional complex vector space $\mathbb{C}^{2}$,
\begin{equation}
  \label{eq:HAB4RUZ8E}
\mathbf{z}=
\begin{pmatrix}z_{1}\\z_{2}\end{pmatrix}
  \mbox{,}
  \end{equation}
  with inner product $(\mathbf{z},\mathbf{w})$ defined according to
  $(\mathbf{z},\mathbf{w})=\mathbf{z}^{+}\mathbf{w}=z_{1}^{*}w_{1}+z_{2}^{*}w_{2}$.  The norm
  $\|\mathbf{z}\|$  of $\mathbf{z}$ is defined by  $\|\mathbf{z}\|^{2}=(\mathbf{z},\mathbf{z})$.
  Writing $z_{1}=a+\imath b,\,z_{2}=c+\imath d$ the points of  $\mathbb{C}^{2}$ may be used to
  assign coordinates $a,b,c,d$ to the points of $\mathbb{R}^{4}$.  The $3-$sphere $S_{3}$ is the
  subset of $\mathbb{R}^{4}$ defined by $\|\mathbf{z}\|^{2}= a^{2}+b^{2}+c^{2}+d^{2}=1$,
  \begin{equation}
    \label{eq:TRU2ZWH87J}
S_{3}=\left\{\mathbf{z} \in \mathbb{C}^{2}| \,\|\mathbf{z}\|^{2}=1\right\}
    \mbox{.}
    \end{equation}
In order to characterize the topological (homotopy) class of the mapping from $\mathbb{R}^{3}$ to our
two-dimensional manifold $Q,\,P$, the subset $S_{3}$  may be identified with $\mathbb{R}^{3}$
(it is in fact a compactified version of $\mathbb{R}^{3}$~\cite{naber:topology}). The Hopf map
is a many to one map from $S_{3}$ to the $2-$sphere $S_{2}$, defined by
\begin{equation}
  \label{eq:DID3DO7HP}
h^{i}(x)=z_{\alpha}^{*}\sigma^{i}_{\alpha\beta}z_{\beta}
  \mbox{,}
  \end{equation}
  where $x$ stands for $a,b,c,d$, the indices $\alpha$ and $\beta$ run from $1$ to $2$, and $\sigma^{i}$
  are the Pauli matrices
  \begin{equation}
    \label{eq:GLI5DP7M1N}
\sigma^{1}=\left(\begin{array}{ccc}0 & 1\\1 & 0\end{array}\right), \;\;\;
\sigma^{2}=\left(\begin{array}{ccc}0 & -\imath \\\imath & 0\end{array}\right), \;\;\;
\sigma^{3}=\left( \begin{array}{ccc}1 & 0 \\0 & -1\end{array}\right)
     \mbox{.}
    \end{equation}
In order to verify this we use the formula
$\sigma^{i}_{\alpha\beta} \sigma^{i}_{\gamma\delta}=2\delta_{\alpha\delta}\delta_{\gamma\beta} - \delta_{\alpha\beta}\delta_{\gamma\delta}$ to obtain $h^{i}h^{i}=\left(|z_{1}|^{2}+ |z_{2}|^{2}\right)^{2}$. Thus, the image of the Hopf
map is indeed a  point of $S_{2}$ given that $|z_{1}|^{2}+ |z_{2}|^{2}=a^{2}+b^{2}+c^{2}+d^{2}=1$.
The points of   $S_{2}$ have coordinates
\begin{gather}
  \label{eq:HDJU7GHR7N}
h^{1}=2(ac+bd)
\mbox{,}\\
\label{eq:HD12AWEEN}
h^{2}=2(ad-bc)
\mbox{,}\\
  \label{eq:HMO4MIMEN}
h^{3}=a^{2}+b^{2}- c^{2}- d^{2}
\mbox{.}
\end{gather}  
Two points $\mathbf{z},\,\mathbf{w}$ in $S_{3}$, which differ from each other by a complex number of
magnitude 1, are mapped to the same point of $S_{2}$.  The set of all points $\mathrm{e}^{\imath \chi}\mathbf{u}$
on $S_{3}$ that results from a fixed $\mathbf{u}$  is called a fiber. Each fiber is the preimage of a point
of $S_{2}$. Thus $S_{3}$ has the structure of a fiber bundle.

The points on $S_{3}$ can now be represented in the form
\begin{equation}
  \label{eq:T5WERZ3GNZ}
  \mathbf{z}=
  \mathrm{e}^{\imath \frac{\chi}{2}}
  \begin{pmatrix}a+\imath b\\c+\imath d\end{pmatrix},\;\;\;\;
a^{2}+b^{2}+c^{2}+d^{2}=1
\mbox{.}
  \end{equation}
 It will become clear soon why it is convenient to use the definition $\frac{\chi}{2}$ in the prefactor.
  It makes sense to introduce new variables that eliminate the surface constraint
  in~\eqref{eq:T5WERZ3GNZ}. We use spherical coordinates, where the angle $\varphi$
  is measured from the $y$-axis in clockwise direction, to represent the three-vector $\mathbf{h}$ in
  the form
 \begin{equation}
   \label{eq:DREI23KL8NL}
   \mathbf{h}=\sin\vartheta  \sin\varphi\, \mathbf{e_{1}}+
   \sin \vartheta \cos \varphi \, \mathbf{e_{2}}+
   \cos \vartheta\,  \mathbf{e_{3}}
   \mbox{.}
   \end{equation}
 Using now Eqs.~\eqref{eq:HDJU7GHR7N}-\eqref{eq:HMO4MIMEN} the dependence of
 $a,b,c,d$ on $\vartheta,\,\varphi$ may be determined and we obtain
 \begin{equation}
   \label{eq:HALE2DKR9WG}
   \mathbf{z}=
   \begin{pmatrix}z_{1}\\[0.15cm]z_{2}\end{pmatrix}=
\mathrm{e}^{\imath \frac{\chi}{2}}
   \begin{pmatrix}u_{1}\\[0.15cm]u_{2}\end{pmatrix}=
 \mathrm{e}^{\imath \frac{\chi}{2}}
 \begin{pmatrix}\cos \frac{\vartheta}{2}\mathrm{e}^{\imath \frac{\varphi}{2}}\\[0.15cm]
\imath \sin \frac{\vartheta}{2}\mathrm{e}^{-\imath \frac{\varphi}{2}}
  \end{pmatrix}
   \mbox{,}
   \end{equation}
 as a possible representation of $\mathbf{z}$. In the literature, essentially two different but equivalent
 representations are in use, which are due to Takabayasi~\cite{takabayasi:description} and
 Bohm~\cite{bohm_schiller_tiomno:causal}, respectively. Both were introduced in order to rewrite Pauli's
 equation in “hydrodynamic form”. The state vector derived here is associated with the description of rotations
 in terms of Euler angles and agrees with Bohm's representation; it was already used in Ref.~\cite{klein:nonrelativistic} ,

 In order to better understand the physical meaning of the quantity $\chi$, we introduce a three-vector
 $\mathbf{t}$, with components defined by $t^{i}=\{\mathbf{z},\sigma^{i}\mathbf{z}\}$ . The curly brackets
 denote here an antisymmetric product, defined by $\{\mathbf{z},\mathbf{w}\}=z_{1}w_{2}-z_{2}w_{1}$.
 This vector fulfills the relations  $\mathbf{t}^{2}=1$ and $\mathbf{t}\mathbf{h}=0$, i.e. $\mathbf{t}$, as a 
unit vector perpendicular to $\mathbf{h}$, lies in the tangential plane determined by $\mathbf{h}$.  
Using the representation~\eqref{eq:HALE2DKR9WG} we obtain $\mathbf{t}$ in terms of the angles
$\chi,\,\vartheta,\,\varphi$:
\begin{equation}
   \label{eq:TRPE8KJU25G}
   \begin{pmatrix}t^{1}\\t^{2}\\t^{3}
   \end{pmatrix}
   =
    \begin{pmatrix}\cos\chi \cos\varphi-\sin\chi  \cos\vartheta   \sin\varphi \\
      -\cos\chi \sin \varphi-\sin\chi \cos\vartheta   \cos\varphi\\
      \sin\chi  \sin\vartheta 
  \end{pmatrix}
   \mbox{.}
   \end{equation}
This relation shows that $\chi$  may be interpreted as an angle of rotation around an axis determined by
$\vartheta$  and $\varphi$. Remarkably, the rotation angle of the vector $\mathbf{t}$ is  $2\alpha$ if
the phase of $\mathbf{z}$ changes by $\alpha$. In particular, $\mathbf{z}$ changes its sign when
$\chi$ changes by $2 \pi $ and returns to its original value  only when $\chi$ changes by $4 \pi $. An  equivalent
form of a “spinor”, as a directed quantity that describes a rotation, was derived by Payne using intuitive geometric
methods\cite{payne:spinor}. 

\subsection[Helicity and Hopf invariant]{Helicity and Hopf invariant}
\label{sec:helic-hopf-invar}
The preimage of each point $P$ of  $S_{2}$ is a circle. The topological nontriviality of the Hopf map is
given by the fact that the linking number for every $2$  circles, that are mapped to different points
$P_{1},\,P_{2}$, is not $0$  but $1$ ; an explicit proof may e.g. be found in~\cite{lyons:hopf}.
This linking number is a topological invariant referred to as Hopf invariant $\gamma$ .
Whitehead~\cite{whitehead:hopf} found that $\gamma$ may  be written as an integral, 
\begin{equation}
  \label{eq:DH3INV8CNB}
\gamma=\frac{1}{16\pi^{2}}\int\,\mathrm{d}^{3}q\,C_{i}\,D_{i}
  \mbox{,}
  \end{equation}
  where $C$, the canonical connection of the Hopf map~\cite{urbantke:hopf}, is given by
  $C_{i}=-2\imath z_{a}^{*}\partial_{i}z_{a}$, and $D_{i}=\epsilon_{ijk}\partial_{j}C_{k}$.   
  The value of the Hopf invariant $\gamma$  is a property of the divergence-free field $D$.
 The field $C$ plays the role of a vector potential for $D$  and is not gauge-invariant. Conversely,
 one may choose any suitable $C$ from the class leading to $D$ in order to calculate $\gamma$,
 as will be done here.

 The basic fields $\chi,\,\vartheta,\,\varphi$ that determine~\eqref{eq:DH3INV8CNB} are
 dimensionless quantities while  the fields $C$  and $D$  have dimensions $cm^{-1}$ and $cm^{-2}$.
 If one wants to use the Whitehead integral in physics, the  abstract fields $C$  and $D$ have to  be
 replaced by suitable physical fields  (e.g. velocity and vorticity, or vector potential and magnetic field),
 and appropriate dimensions and pre-factors must be introduced. In our case we have to establish the
 relation between the field $C$ and the momentum field $M$,
 \begin{equation}
   \label{eq:HABA2EJU9Z}
   C_{i}=\partial_{i }\chi+\cos\vartheta\partial_{i}\varphi\;\;\;\;\;\;\;\;
   M_{i}=\partial_{i }S+P\partial_{i}Q
   \mbox{,}
   \end{equation}
in order to be able to find the relation between $S,\,P,\,Q$ and $\chi,\,\vartheta,\,\varphi$.  If we accept
$\mathbf{z}$, as defined by~\eqref{eq:HALE2DKR9WG}, as our new state variable (apart from an amplitude
which we will be introduced later) then it is obvious to associate the two-component quantity 
in~\eqref{eq:HALE2DKR9WG}, with components $u_{1}, u_{2}$, with the new rotational degrees of
freedom $Q,P$. As a consequence we identify the prefactor $\mathrm{e}^{\imath \frac{\chi}{2}}$ with the
earlier phase factor $\mathrm{e}^{\imath \frac{S}{\hbar}}$, associated with the irrotational momentum fields
studied in III. This leads to the relations
\begin{equation}
  \label{eq:HEN7DL13GGF}
  \frac{S}{\hbar}=\frac{\chi}{2},\;\;\;C_{i}=\frac{2}{\hbar}M_{i},\;\;\;P\partial_{i}Q=
  \frac{\hbar}{2}\cos\vartheta\partial_{i}\varphi
  \mbox{,}
  \end{equation}
 As for the identification of $P,Q$ we could, remembering that the Clebsch potentials are canonical
 variables~\cite{rund:clebsch}, interpret $P$ as momentum and $Q$ as position. This could be achieved
 by replacíng $\hbar / 2$ by $\hbar / 2R$ and $\phi$ by $R\phi$ , where $R$ is a length. The numerical
 value of $R$ is arbitrary since $P, Q$ always occur in pairs. Thus we may set $R=1$:
 \begin{equation}
   \label{eq:HQR3TUZ87DF}
P=\frac{\hbar}{2}\cos\vartheta,\;\;\;Q=\varphi
   \mbox{.}
 \end{equation}
 The topological meaning of the ``canonical Clebsch potentials'' defined by~\eqref{eq:HQR3TUZ87DF},
 was clarified by Kutnetsov and Mikhailov~\cite{kuznetsov:topological}. These authors studied ideal fluids
 which are, however, described by essentially the same mathematical structure as the present problem. In
 their work the constant $\hbar/2$ is replaced by an undetermined constant - let us recall that we were
 only able to fix the value of this constant by anticipating the quantization process.

 If we replace $P,Q$ in~\eqref{eq:HAPR23KR4D2F} by the canonical Clebsch potentials~\eqref{eq:HQR3TUZ87DF}
  the vorticity $\Omega_{i}$  and the helicity $H_{\Omega}$ become quantized variables,     
   \begin{equation}
\label{eq:RUZR12PE4D2F}
\Omega_{i}= \frac{\hbar}{2}\epsilon_{ikl} \left( \partial_{k}\cos\vartheta\right)\left(\partial_{l} \varphi\right),
\;\;\;\;\;\;\;
H_{\Omega}=(2\pi\hbar)^{2}\gamma
\mbox{.}
\end{equation}
The linking number $\gamma$ takes integer values in general, and is $1$ in the present theory.   
The Hopf map is sometimes introduced by restricting the vorticity according to the relation
$\Omega_{i}= \frac{\hbar}{2}T_{i}$, where 
\begin{equation}
  \label{eq:D24LRVQSTT}
T_{i}=\frac{1}{2}\epsilon_{ijk}\epsilon_{lmn}h^{l}\left(\partial_{j}h^{m} \right)\left(\partial_{k}h^{n} \right)  
  \mbox{.}
  \end{equation}
The vector $\mathbf{T}$  was first introduced by Takabayasi~\cite{takabayasi:vector} and later
rediscovered in other contexts by Faddeev~\cite{faddeev:solitons} and Mermin and
Ho~\cite{mermin_ho:circulation}. Topological methods have been used successfully to classify phases
in superfluid He and other many-body systems~\cite{mermin:topological}. If $\mathbf{v}$  in
Eq.~\eqref{eq:DE4GHL9MMA} is replaced by the superfluid velocity~\cite{volovik_mineev:particle}
one obtains the same $\hbar$-dependent prefactor as in Eq.~\eqref{eq:RUZR12PE4D2F}. 

\subsection[Invariance of circulation and quantization condition]{Invariance of circulation and quantization condition}
\label{sec:invar-circ-quant}
As shown in III,  after projection to the $n$-dimensional subspace defined by $M$, the Poincar\'{e} integral
invariant takes the form
\begin{equation}
  \label{eq:HZGR7MSU5LA}
I_{\bar{C}}(t)=\oint_{\bar{C}_{t}}\,M_{i}(q,t)\mathrm{d}q_{i}
\mbox{,}
\end{equation}
where $\bar{C}_{t}$ is a closed path in the subset $(q,M(q,t))$ of phase space. This formula expresses
the invariance of the circulation $I_{\bar{C}}$;  in a  fluid-dynamical context it is referred to as Kelvin's theorem.
It remains true if $M$ is expressed in terms of $S,Q,P$ according to~\eqref{eq:REP23HZU3DF}, or in
terms of $S,\vartheta,\varphi$ according to~\eqref{eq:HEN7DL13GGF}.

We expect, in analogy to the helicity, that the invariant $I_{\bar{C}}$ becomes ``quantized'' for topological
reasons. This is indeed the case as found already by Takabayasi, who used - contrary to the present work - the
basic equations of QT as his starting point\cite{takabayasi:vortex}.
Using~\eqref{eq:HABA2EJU9Z}-\eqref{eq:D24LRVQSTT} the circulation may be written as
\begin{equation}
  \label{eq:HZ2TGRPT8LA}
\oint_{\bar{C}_{t}}\,M_{i}(q,t)\mathrm{d}q_{i}=h\left(n+\frac{1}{4\pi}\int_{S_{t}}\mathrm{d}S_{i}T_{i} \right)
\mbox{,}
\end{equation}
where $S_{t}$ is a surface [a cross section of the tube formed by the solutions of~\eqref{eq:UMW29ISJG}]
with boundary $\bar{C}_{t}$. The first term on the r.h.s. is due to the fact, that $S$ may change by
$2\pi\hbar n$ (where $n$  is an integer) when going around  $\bar{C}_{t}$. This multi-valuedness, and its  associated
topological singularity, does not affect the uniqueness of the state function $\mathbf{z}$. 
This term corresponds to the usual quantization condition. The second term on the r.h.s. is due to the
additional non-singular vorticity of the momentum field. 

To interpret this second term, we note that the vortical part of the momentum field may be
interpreted, apart from a constant of proportionality, as a kind of internal vector potential,
with components $A_{k}^{I}(q,t)$. When expressed in terms of  the canonical Clebsch potentials,
$A_{k}^{I}$ is given by
\begin{equation}
  \label{eq:LHRE8BNB13W}
  A_{k}^{I}=-\frac{\hbar c}{2e} \cos\vartheta \partial_{k} \varphi
  \mbox{.}
  \end{equation}
The associated inner magnetic field is proportional to the Takabayasi vector  or to the vorticity,
$B_{i}^{I}=\epsilon_{ikl}\partial_{k}A_{l}^{I}=\frac{c}{e}\Omega_{i}$. It is useful to introduce, for comparison,
an external vector potential with associated magnetic field $B_{i}^{E}$ (This does not cover the complete
influence of the external magnetic field, the full theory will be given in section~\ref{sec:intr-gauge-field}).
The circulation extended this way may be written as
\begin{equation}
  \label{eq:AUE8TGRHT3OA}
  \oint_{\bar{C}_{t}}\,M_{i}(q,t)\mathrm{d}q_{i}
+ \frac{e}{c}\int_{S_{t}}\mathrm{d}S_{i}\left(B_{i}^{E}+B_{i}^{I} \right)
  =n h
\mbox{.}
\end{equation}
We see that the vortical part of the momentum field may be interpreted as an additional,
``internal''  contribution to the magnetic field. In superconducting many-particle systems one observes
the phenomenon of flux quantization, which is described by Eq.~\eqref{eq:AUE8TGRHT3OA} \emph{without}
this vortical contribution. Its absence is, however, not really unexpected, since the superconducting state
is not generated by single electrons but by spinless paired electrons (Cooper pairs).

\section[Semilinear evolution equation]{Semilinear evolution equation}
\label{sec:semil-evol-equat}
After introducing the canonical Clebsch variables, one obtains the updated set of basic equations for the
dynamical variables $\rho, S, \vartheta, \varphi$ by replacing $P,\,Q$ in
Eqs.~\eqref{eq:LBRCS2CEW}-\eqref{eq:T39AH4Q3UZ} by $\cos\vartheta,\,\varphi$ according
to~\eqref{eq:HQR3TUZ87DF}:
\begin{gather}
\partial_{t} S+e\Phi+ 
\frac{\hbar}{2}\cos\vartheta\, \partial_{t} \varphi+\frac{1}{2m}\sum_{k}
\left(\partial_{k}S+ \frac{\hbar}{2}\cos\vartheta\, \partial_{k} \varphi\right)^{2}=  0
\mbox{,}
  \label{eq:LBRGR9WEEW}\\
  \left[\partial_{t}+\left(\partial_{k}S + \frac{\hbar}{2}\cos\vartheta\,\partial_{k}\,\varphi \right)\partial_{k} \right]
  \varphi=0
\mbox{,}
\label{eq:E7ESI3UZAQ}\\
\left[\partial_{t}+\left(\partial_{k}S + \frac{\hbar}{2}\cos\vartheta\,\partial_{k}\,\varphi \right)\partial_{k} \right]
\vartheta=0
\mbox{,}
\label{eq:T89ADU7I9Q}\\
\partial_{t} \rho+\partial_{k}\rho
\left(\partial_{k}S + \frac{\hbar}{2}\cos\vartheta\,\partial_{k}\,\varphi \right)=0
\label{eq:THABIOZ42UZ}
\mbox{.}
\end{gather}
Instead of the two dynamical equations for $\vartheta$ and $\varphi$  the relation
$D_{t}\mathbf{h}=0$ for the unit vector $\mathbf{h}$, defined by Eq.~\eqref{eq:DREI23KL8NL}, may
equivalently be used. The particle equations of motion $\dot{q}_{k}=v_{k}(q,t)$ remain true; they define
the ranges of integration in the quantization condition~\eqref{eq:AUE8TGRHT3OA}.

The above evolution equations, as given by~\eqref{eq:LBRGR9WEEW}-~\eqref{eq:THABIOZ42UZ},
are neither suitable for the transition to QT by linearization, nor for the introduction of a gauge field.
To achieve these goals, these four equations must first be replaced by a single equation for a single
variable with two complex components. A suitable quantity is provided by the Hopf map. 
In order to perform this transformation, it is convenient to
rewrite~\eqref{eq:LBRGR9WEEW}-\eqref{eq:THABIOZ42UZ} in the following form:
\begin{gather}
D_{t}S= \bar{L} 
\mbox{,}
\label{eq:UHTR8EDW}\\
D_{t}\vartheta= 0
\mbox{,}
\label{eq:UHAL3UICW}\\
D_{t}\varphi= 0
\mbox{,}
\label{eq:UHP24O5IW}\\
D_{t}\rho^{\frac{1}{2}}=-\frac{1}{2}\rho^{\frac{1}{2}}\partial_{k}v_{k}
\label{eq:UE8GHH3XB}
\mbox{,}
\end{gather}
where $D_{t}=\partial_{t}+v_{k}\partial_{k}$, $\bar{L}=\frac{1}{2}v_{k}M_{k}-V$, and
\begin{equation}
  \label{eq:DH21RLGW9Q}
v_{k}=\frac{1}{m}\left(\partial_{k}S+\frac{\hbar}{2}\cos\vartheta \,\partial_{k}\varphi\right)
  \mbox{.}
  \end{equation}
The basic equations of QA reformulated this way differ from the equations used in III only with
regard to the additional equations of motion~\eqref{eq:UHAL3UICW} and~\eqref{eq:UHP24O5IW}
and the additional vortical term in the definition~\eqref{eq:DH21RLGW9Q}. The structural
differences between these equations and the original phase space equations used in I, II were
mentioned in III.

Let us stress once again, that these basic equations do essentially not belong to QT, despite the
occurrence of the constant $\hbar$. The occurrence of $\hbar$ is explained as follows: Due
to topological considerations, the introduction of a quantity with the dimension of an action became
necessary. For the sake of simplicity, we have already assigned the special numerical value of Planck's
constant to this quantity; without this assignement the transition to QT, to be performed later,
cannot be realized. This assignment of a numerical value may therefore be interpreted as a first step in
the quantization process. However, the very occurence of this constant is a consequence of classical
(topological) considerations. 

We obtain an appropriate  two-component variable $\psi$, suitable for linearization, by multiplying
the variable $\mathbf{z}$  defined by the Hopf map [see Eq.~\eqref{eq:HALE2DKR9WG}] by the
factor $\sqrt{\rho}$,
\begin{equation}
  \label{eq:DH92RLK2VP}
  \psi_{a}=\sqrt{\rho}\,\mathrm{e}^{\frac{\imath}{\hbar}S}\,u_{a},\;\;\;a=1,2
  \mbox{.}
\end{equation}
Then, as a consequence of the physical meaning of $\rho$, the integral of $|\psi|^{2}$ over the
entire space must be $1$.

Let us write the projected Liouville equation~\eqref{eq:UE8GHH3XB} in the form $\mathbb{L}\rho^{\frac{1}{2}}=0$,
with the differential operator $\mathbb{L}=\frac{\hbar}{\imath}\left[D_{t}+\frac{1}{2}\left(\partial_{i}v_{i} \right) \right]$.  
Let us next consider the two-component quantity  $T_{a}=\mathbb{L}\sigma^{0}_{ab}\psi_{b}$. The terms in
the differential equation to be constructed are necessarily $2 \times 2$ matrices. It turns out that it is sufficient
to use the $2 \times 2$ identity matrix $\sigma^{0}$, at least in the present field-free case. If we let the
differential operator $\mathbb{L}$ act on $\psi$, then  $T_{a}$ takes the form
\begin{equation}
  \label{eq:CR25BSM9AR}
  T_{a}=\frac{\hbar}{\imath}\sigma^{0}_{ab}\psi_{b}
  \left[\rho^{-\frac{1}{2}}\left(D_{t}+\frac{1}{2}\left(\partial_{i}v_{i} \right)\right)\rho^{\frac{1}{2}}+\frac{\imath}{\hbar}D_{t}S+\frac{1}{u_{(b)}}D_{t}u_{(b)}\right]
  \mbox{,}
  \end{equation}
whithout summation over $b$ in the bracket. If we now use the evolution
equations~\eqref{eq:UHTR8EDW}-\eqref{eq:UE8GHH3XB}, the first and third terms vanish
and $D_{t}S$ is replaced by $\bar{L}$. If we equate the resulting expression with the original definition
of $T_{a}$, we obtain the differential equation
\begin{equation}
  \label{eq:WEO2TTHE49D}
  \left[\frac{\hbar}{\imath}D_{t}+\frac{\hbar}{2\imath}\left(\partial_{i}v_{i} \right)-\frac{1}{2}v_{k}M_{k}+V \right]
  \psi_{a}=0
  \mbox{,}
  \end{equation}
which is equivalent to Eqs.~\eqref{eq:UHTR8EDW}- \eqref{eq:UE8GHH3XB}, but is already more
similar to the desired form. In order to proceed with the linearization we use the relation
$\left(\partial_{k}S\right)\psi_{b}=\frac{\hbar}{\imath}\left[\partial_{k}-\rho^{-\frac{1}{2}}\left(\partial_{k} \rho^{\frac{1}{2}}\right)+u_{(b)}^{-1}\left(\partial_{k}u_{(b)}\right)\right]\psi_{b}$ which follows directly from the definition of $\psi$
[see Eq.~\eqref{eq:DH92RLK2VP}]. Using this last relation as well as the definition of $M_{k}$ one obtains
the equation $v_{i}M_{i}\psi_{b}=\frac{\hbar}{\imath}v_{i}\left(\partial_{i}-f_{i(b)}\right)\psi_{b}$. The function
\begin{equation}
  \label{eq:DE5O1U9ZWQ}
  f_{kb}=\frac{1}{\rho^{\frac{1}{2}}}\left(\partial_{k} \rho^{\frac{1}{2}}\right)+
  \frac{1}{u_{(b)}}\left(\partial_{k}u_{(b)}\right)-\frac{\imath}{2}\cos\vartheta \,\left( \partial_{k}\varphi \right)
  \mbox{,}
  \end{equation}
does not depend on the variable $S$ which carries the gauge degree of freedom. Repeated application of
this last equation leads, after a number of further elementary rearrangements, to the following ``semilinear''
differential equation
\begin{equation}
  \label{eq:WRO76TDE49Q}
  \left[\frac{\hbar}{\imath}\partial_{t}-\frac{\hbar^{2}}{2m}\partial_{k}\partial_{k}+V \right]\psi_{a}=
  -\frac{\hbar^{2}}{2m}
  \left[f_{k(a)}f_{k(a)}+(\partial_{k}f_{k(a)}) \right]\psi_{a}
  \mbox{.}
  \end{equation}
 This differential equation, together with the particle equation of motion $\dot{q}_{k}=v_{k}(q,t)$,
 represents that form of the QA which is most similar to QT - and which accordingly enables a
 particularly simple transition to QT. In I a differential equation in phase space was derived (Eq. 19 of I),
 which has been called the “classical counterpart of Schrödinger's equation”. If we follow this naming scheme,
 we can call Eq.~\eqref{eq:WRO76TDE49Q}  the “quasiclassical counterpart of Schrödinger's equation”. 
 It represents the \emph{completion} of the non-linear Schr\"odinger equation reported in III (see Eq. 40
 of III), where two of the three possible degrees of freedom of the momentum field were unjustifiably
 neglected.

\section[Introducing a gauge field]{Introducing a gauge field}
\label{sec:intr-gauge-field}
Let us now “switch on” an electromagnetic field. This is usually done using the principle
of minimal coupling
\begin{equation}
  \label{eq:HP28OI57WV}
  \frac{\hbar}{\imath}
  \partial_{k} \Rightarrow
  \frac{\hbar}{\imath}\partial_{k}-\frac{e}{c}A_{k},\;\;\;\;\;\;
  -\frac{\hbar}{\imath} \partial_{t} \Rightarrow
  -\frac{\hbar}{\imath} \partial_{t}+e\Phi 
  \mbox{,}
  \end{equation}
where $A,\,\Phi$ is the vector and scalar potential respectively. In the (quasi) classical
equation~\eqref{eq:WRO76TDE49Q} application of this rule to the derivatives on the right-hand side
does not lead to meaningful results. We therefore use the more fundamental version of the principle
of minimal coupling formulated by Dirac~\cite{dirac:quantised} (see also~\cite{kaempfer:concepts},~\cite{klein:schroedingers}). With this method, the wave function $\psi$ is multiplied by a non-integrable phase factor:
\begin{equation}
  \label{eq:NIPH13FLIS7M}
  \exp{\bigg\{
    -\frac{\imath}{\hbar}\frac{e}{c}
    \int^{x,t}\left[\mathrm{d}q_{k}' A_k(q',t')-c\mathrm{d}t' \Phi(q',t')\right]
 \bigg\}}
  \mbox{,}
  \end{equation}
This phase factor is then shifted to the left of the differential operators, creating
the potentials, and can afterwards be eliminated. The final wave function is again single-valued as
it should be. This version of the minimal coupling rule can be applied to
Eq.~\eqref{eq:WRO76TDE49Q}, since the right-hand side does not contain any
derivatives of $\psi$ . We obtain in this way the standard result
\begin{equation}
  \label{eq:WKL35THT49Q}
  \left[\frac{\hbar}{\imath}\partial_{t}-e\Phi-\frac{\hbar^{2}}{2m}
 \sum_{k=1}^{3} \, \left( \partial_{k}-\imath\frac{e}{\hbar c}A_{k}\right)^{2}
    +V \right]\psi_{a}=
  -\frac{\hbar^{2}}{2m}
  \left[f_{k(a)}f_{k(a)}+(\partial_{k}f_{k(a)}) \right]\psi_{a}
  \mbox{.}
  \end{equation}
The linear part of Eqs.~\eqref{eq:WRO76TDE49Q} and~\eqref{eq:WKL35THT49Q}
apparently plays a decisive role in the transition to QT. This part, in which, based on our
assumptions, the kinetic energy is proportional to the identity matrix, does not contain any
coupling between the two components of $\psi$, neither in the field-free case nor in the presence
of an electromagnetic field. Assuming that Eq.~\eqref{eq:WRO76TDE49Q} is true, this would
lead to the strange conclusion that the vortical components of the momentum field have no observable
consequences at all after transition to QT. Instead, we suspect that Eq.~\eqref{eq:WRO76TDE49Q}
is incomplete. 

\subsection[Two representations of Euclidean space]{Two representations of Euclidean space}
\label{sec:two-repr-eucl}
The geometric objects occurring in physics may be characterized by their behavior under certain
groups of transformations. The relevant group in the present context is the rotation group $SO(3)$,
the group of orthogonal $3 \times 3$ matrices with determinant $1$. A geometric object is associated
with a specific representation of $SO(3)$. For example, in the Schr\"odinger equation for spinless
particles studied in III, a one-component complex quantity occurs as a geometric object. In this case,
only the space of independent variables is transformed under $SO(3)$ and the representation
associated with this geometric object is the identical representation.

The geometric object that arises in the present problem is the spinor $\psi$.
At every point of space this is an element of a Hilbert space  $\mathbb{H}^{2}$, which is
defined (as in section~\ref{sec:hopf-map}) by  $\mathbb{C}^{2}$ with the inner product
$(\psi,\phi)=\psi_{1}^{*}\phi_{1}+\psi_{2}^{*}\phi_{2}$. The group of automorphisms of
$\mathbb{H}^{2}$ is given by the linear transformations that leave the inner product invariant.
This is the group $U(2)$, but it can be restricted to $SU(2)$, the group of two-dimensional unitary
transformations with determinant $1$. A general element $U \in SU(2)$ has the form
\begin{equation}
  \label{eq:AG1EO3SU28H}
U=\left(\begin{array}{ccc}\alpha & \beta\\-\beta^{*} &\alpha^{*} \end{array}\right)
  \mbox{,}
  \end{equation}
where $\alpha$ and $\beta$ are complex numbers that satisfy $|\alpha|^{2}+|\beta|^{2}=1$ . This means
that the group space of $SU(2)$ is given by the $3$-sphere $S_{3}$ (the spinorial part of $\psi$ is also
determined by a point on $S_{3}$, as was shown in section~\ref{sec:hopf-map}). Thus, a spinor is the
geometric object transforming according to the natural representation of the group $SU(2) \cong S_{3}$.

Let us recall the relation between the groups $SU(2)$ and $SO(3)$. This point  is well-known
from the quantum mechanical theory of spin; but of course this relation is primarily a “classical”
matter of group theory and associated topological considerations. To find this relation we introduce
a matrix representation of Euclidean space following Eberlein~\cite{eberlein:spin_model}. We start
from the observation that every Hermitian operator in $\mathbb{H}^{2}$ may be represented as a
linear combination, with real coefficients, of the three Pauli matrices $\sigma^{k}$  and the
identity matrix $\sigma^{0}$ . If we associate in our non-relativistic theory the identity matrix
$\sigma^{0}$ with a time-like coordinate, then the remaining three Pauli matrices $\sigma^{k}$,
with trace $0$, may be used to represent an arbitrary three-dimensional Hermitian $2 \times 2$
matrix, say $Q$, in the form
\begin{equation}
  \label{eq:DF28ZHZ9HRI}
Q=q_{1}\sigma^{1}+q_{2}\sigma^{2}+q_{3}\sigma^{3}
  \mbox{,}
  \end{equation}
where the $q_{i}$ are real numbers. Let us denote the Euclidean space with the usual inner product
by $E_{3}$ and  the real vector space of the Hermitian matrices~\eqref{eq:DF28ZHZ9HRI}, with basis
elements $\sigma^{k}$, by $\mathbb{E}_{3}$.  There is a one-to-one correspondence between the
vectors $q_{i}\mathbf{e}_{i} \in E_{3}$, where $\mathbf{e}_{i}$ is an orthonormal basis, and the
matrices $q_{k}\sigma^{k} \in \mathbb{E}_{3}$,
\begin{equation} 
  \label{eq:ALLK9BREI3U}
q_{i}\mathbf{e}_{i} \leftrightarrow q_{k}\sigma^{k}
  \mbox{.}
  \end{equation}
This mapping from $E_{3}$ to $\mathbb{E}_{3}$ also preserves the inner product $(p,q)=p_{i}q_{i}$. If $P$
and $Q$ are the elements of $\mathbb{E}_{3}$ corresponding to $p$ and $q$, then their inner product,
which is denoted by $(P \cdot Q)$, is defined by $PQ+QP=2(P \cdot Q)\sigma^{0}$.    
The invariance can easily be verified with the help of the relation
$\sigma^{i}\sigma^{k}+\sigma^{k}\sigma^{i}=2 \delta^{ik} \sigma^{0}$ . Thus, the space $\mathbb{E}_{3}$ 
may be called the matrix model, or the spin model, of Euclidean
space~\cite{naber:topology},\cite{eberlein:spin_model}. 

Let us consider an element $Q \in \mathbb{E}_{3}$ and a unitary matrix $U \in SU(2)$. The transformed
element
\begin{equation}
  \label{eq:ATREI4LV8M}
R_{U}(Q)=UQU^{-1}
  \mbox{}
  \end{equation}
is hermitian and has vanishing trace, so it is again an element of $\mathbb{E}_{3}$ which
may be written in the form $Q^{'}=q^{'}_{k}\sigma^{k}$. If we now insert $U$ as given
by~\eqref{eq:AG1EO3SU28H} and evaluate Eq.~\eqref{eq:ATREI4LV8M} we find the relation
\begin{equation}
  \label{eq:WFA5WRN8BI}
q^{'}_{i}=R_{ik}(\alpha,\beta)q_{k}
  \mbox{,}
  \end{equation}
where $R_{ik}(\alpha,\beta)$ are the elements of a (real) orthogonal matrix with determinant $1$ which is uniquely
determined by the matrix $U$ defined by~\eqref{eq:AG1EO3SU28H}. These elements satisfy the relationship
$R_{ik}(-\alpha,-\beta)=R_{ik}(\alpha,\beta)$. As a consequence, Eq.~\eqref{eq:ATREI4LV8M} defines a
mapping $SU(2) \rightarrow SO(3)$ which is not one-to-one, since the two elements $\pm U$ are
mapped onto the same element of $SO(3)$. One can also show that $R_{U_{1}U_{2}}=R_{U_{1}}R_{U_{2}}$
and that every rotation in $\mathbb{E}_{3}$ may be represented in the
form~\eqref{eq:ATREI4LV8M}~\cite{eberlein:spin_model},\cite{naber:topology}.
The mapping $SU(2) \rightarrow SO(3)$, which is often referred to as ``spinor mapping'', is therefore
a two-to-one group homomorphism.

The difference between the groups $SU(2)$ and $SO(3)$ is essentially due to the different
topological structure of their parameter spaces. The two-to-one group homomorphism from
the simply connected $SU(2)$ to the double connected $SO(3)$ means that $SU(2)$ is the
covering group (double cover) of $SO(3)$. As already mentioned, the parameter space of
$SU(2)$ is the $3$-sphere $S_{3}$. The parameter space of $SO(3)$ may be identified with
a filled sphere in three-dimensional space with radius $\pi$; this choice of the group manifold
allows an easy intuitive understanding of the topological structure~\cite{sattinger:lie_groups}.
Alternatively, one may also use the surface of the $3$-sphere $S_{3}$ with antipodal points
identified, as group manifold; this latter choice is particularly plausible in view of the two-to-one
homomorphism from $SU(2)$ to $SO(3)$. 

To choose between the two representations $E_{3}$ and $\mathbb{E}_{3}$ of the abstract
Euclidean space, we note that  the transformation behavior of the Hermitian operator
$Q \in \mathbb{E}_{3}$, as given by~\eqref{eq:ATREI4LV8M}, is consistent with the
general condition that the action of an operator before and after a basis transformation $U$
must be the same. The spin model $\mathbb{E}_{3}$ of Euclidean space is therefore the appropriate
choice for vectors that act as operators on the Hilbert space $\mathbb{H}^{2}$.

The representation $E_{3}$ would be suitable if we had three-vectors as geometric objects, in
which the effect of $SO(3)$ is realized by means of the usual rotation matrices.  The spin group $SU(2)$,
as discussed above, has a richer structure compared to the $SO(3)$. Its group manifold is larger,
as is the number of its representations; any representation of $SO(3)$ can be extended to a
representation of the $SU(2)$ (by composition with the spinor mapping), but not the other way around. 
[However, in the physical literature the term “two-valued representation of $SO(3)$” is often used
for $SU(2)$]. The fact that the simply connected parameter space of $SU(2)$  contains two elements $\pm U$
for every rotation $R$ may be related to the corresponding property of the full group $O(3)$
of the automorphisms of Euclidean space. The statement
\begin{quote}
The orthogonal transformations are the automorphisms of Euclidean vector space. Only with the
spinors do we strike this level in the theory of its representations... 
\end{quote}
made by Hermann Weyl emphasises the fundamental role of the spin
representation~\cite{weyl:classical_groups}.

The spinorial geometric object $\psi$  associated with $SU(2)$ is essentially (except for
a subsequently added amplitude $\sqrt{\rho}$) given by the Hopf map, defined in
section~\ref{sec:hopf-map}. The introduction of the Hopf map was based on the assumption
that the statistical ensembles occurring in nature belong to a topological class describing linked
vortex lines. This choice, which may have seemed a bit arbitrary, is now justified by the fact
that the Hopf map describes the topology of rotations in $\mathbb{H}^{2}$. The property
of linked vortex lines seems to correspond to a particular topological property of rotations, namely the fact
that $\psi$ does not return to the starting point until it has been rotated by $4\pi$. 

\subsection[Minimal coupling in $\mathbb{E}_{3}$]{Minimal coupling in $\mathbb{E}_{3}$}
\label{sec:minim-coupl-mathbbe}
According to the discussion of the last section, the transition from $E_{3}$ to $\mathbb{E}_{3}$
should be performed simultaneously with the introduction of the spinor $\psi$. The transition is defined
by the simple replacement rule $v_{i}\mathbf{e}_{i} \rightarrow v_{i}\sigma^{i}$.
This rule can only be applied in the quantum-like version~\eqref{eq:AG1EO3SU28H}
of the equation of motion, and here again only in the Laplace operator, which can be written in the
form~$\partial_{i}\mathbf{e}_{i}\partial_{k}\mathbf{e}_{k}$ . One obtains
\begin{equation}
  \label{eq:NOZ3EGTU2N}
  \partial_{k}\partial_{k}=
  \partial_{i}\mathbf{e}_{i}\partial_{k}\mathbf{e}_{k}
  \rightarrow
  \partial_{i}\sigma^{i}\partial_{k}\sigma^{k}=
  \partial_{k}\partial_{k}+\imath \epsilon^{lik}
  \sigma^{l} \partial_{i}\partial_{k}
  \mbox{,}
  \end{equation}
where the relation $\sigma^{i}\sigma^{k}=\delta^{ik}+\imath \epsilon^{lik}
\sigma^{l}$ was used to obtain the last term on the right-hand side. Eq.~\eqref{eq:NOZ3EGTU2N}
shows that this replacement rule can only lead to a new contribution, as compared with the original 
Laplace operator, if the first order derivatives are not interchangeable, that is, if they
operate on a a singular, typically multi-valued  $\psi$. This is the case if the minimal
coupling rule~\eqref{eq:NIPH13FLIS7M} is applied. The evolution equation then takes the form
\begin{equation}
  \label{eq:WRZ18THO43Q}
  \left[\frac{\hbar}{\imath}\partial_{t}-e\Phi-\frac{\hbar^{2}}{2m}
 \sum_{k=1}^{3} \, \left( \partial_{k}-\imath\frac{e}{\hbar c}A_{k}\right)^{2}+V\right]\psi_{a}
 +\mu_{B}\sigma_{ab}^{k}B_{k}\psi_{b}=
  -\frac{\hbar^{2}}{2m}
  \left[f_{k(a)}f_{k(a)}+(\partial_{k}f_{k(a)}) \right]\psi_{a}
  \mbox{.}
  \end{equation}
 where $\mu_{B}=-e\hbar/2mc$ is the Bohr magneton, and $B_{k}$ are the components of the
 magnetic field $\mathbf{B}=\mathbf{\nabla} \times \mathbf{A}$. The $B-$dependent coupling term was introduced by
 Pauli and is sometimes referred to as ``Zeemann term'', as it played an important role in the explanation
 of the anomalous Zeeman effect. In mathematical terms $\mathbf{A}$ is called connection and $\mathbf{B}$
 is called curvature of the connection~\cite{naber:topology}.

We brought the quasi-classical evolution Eqs~\eqref{eq:LBRGR9WEEW}-\eqref{eq:THABIOZ42UZ}
in the quantum-like form~\eqref{eq:WRO76TDE49Q} and then introduced an external electromagnetic field
with the help of the principle of minimal coupling. If we now go back from the result~\eqref{eq:BRA39TR56Q}
to the original form, we obtain the equations
\begin{gather}
\partial_{t} S+e\Phi+ 
\frac{\hbar}{2}\cos\vartheta\, \partial_{t} \varphi+\frac{m}{2}\sum_{k}
v_{k}^{2} + \mu_{i}B_{i}=  0
\mbox{,}
  \label{eq:LTRRU9W2EW}\\
  D_{t}\varphi = \frac{e}{mc}\frac{1}{\sin\vartheta}
  \left( B_{3} \sin\vartheta - B_{2}\cos\vartheta \cos\varphi -B_{1}\cos\vartheta \cos\varphi \right) 
\mbox{,}
\label{eq:E7ETR3GAH4}\\
D_{t}\vartheta=\frac{e}{mc} \left( B_{1} \cos\varphi - B_{2}\sin\varphi \right) 
\mbox{,}
\label{eq:T8HABA778Q}\\
\partial_{t} \rho+\partial_{k}\rho v_{k}=0
\label{eq:TTOIUOZ14UZ}
\mbox{,}
\end{gather}
where the velocity field $v$ and the total derivative $D_{t}$ are now defined by
\begin{equation}
  \label{eq:BG2FR3TW8Q}
  v_{k}=\frac{1}{m}\left( \partial_{k}S-\frac{e}{c}A_{k}+ \frac{\hbar}{2}\cos\vartheta\, \partial_{k} \varphi \right),\;\;\;\;
  D_{t}=\partial_{t}+v_{k}\partial_{k}
  \mbox{.}
  \end{equation}
The ``magnetic moment of the electron'' is defined by $\mu_{i}=-\frac{e}{mc}s_{i}$, where $\mathbf{s}$
is a three-vector of constant lenght $\frac{\hbar}{2}$ which is parallel to the unit vector $\mathbf{h}$ defining the
Hopf map,
\begin{equation}
\label{eq:Z8WEH3DFT9R}
\mathbf{s}=\frac{\hbar}{2}
\left(\sin\vartheta  \sin\varphi\, \mathbf{e_{1}}+\sin \vartheta \cos \varphi \, \mathbf{e_{2}}+\cos \vartheta\,  \mathbf{e_{3}} \right)
  \mbox{.}
  \end{equation}
The electromagnetic field leads to two types of new terms in
Eqs.~\eqref{eq:WRZ18THO43Q}-\eqref{eq:TTOIUOZ14UZ}, namely on the one hand
to the usual extensions of the energy and momentum fields by potentials and on the
other hand to terms depending on the magnetic field $\mathbf{B}$. The latter terms may
be interpreted with the help of the vector $\mathbf{s}$, using the fact that
Eqs.~\eqref{eq:E7ETR3GAH4} and~\eqref{eq:T8HABA778Q} may be written in the form
\begin{equation}
  \label{eq:DTG4EWIU3Z}
D_{t}\mathbf{s}=-\frac{e}{mc}\mathbf{B} \times \mathbf{s}
  \mbox{,}
  \end{equation}
which is the field-theoretical version of the equation of motion of a classical magnetic dipole
in a magnetic field.

\subsection[The magnetic moment of the electron]{The magnetic moment of the electron}
\label{sec:magn-moment-elec}
According to classical electrodynamics, a rotating charge distribution with mass $m$, total charge $q$,
and angular momentum $\mathbf{L}$ leads to a magnetic dipole moment
$\boldsymbol{\mu}=\frac{q}{2mc}\mathbf{L}$. The pre-factor $\frac{q}{2mc}$ is called the gyromagnetic ratio.
The interaction with an external magnetic field is taken into account by a term $\boldsymbol{\mu}\mathbf{B}$
in the Hamiltonian function. The new $B-$dependent terms in Eqs.~\eqref{eq:LTRRU9W2EW}
and~\eqref{eq:DTG4EWIU3Z} agree in form with this particle image, differ however with regard
to the gyromagnetic ratio between the “intrinsic angular momentum” $\mathbf{s}$ and the associated
magnetic moment $\boldsymbol{\mu_{s}}$. The gyromagnetic factor $g$ defined by
$\boldsymbol{\mu_{s}}=\frac{g}{2}\frac{q}{mc}\mathbf{s}$ takes the value $g=2$ and is consequently twice as
large as would be expected in the particle image. 

If we want to interpret the Zeemann term in the quantum-like
form~\eqref{eq:WRZ18THO43Q} of the evolution equations, we must first
identify the quantum-like quantity $\mathbf{S}$ corresponding to $\mathbf{s}$. The formula
$\rho \mathbf{s}=\psi^{*}_{a}\frac{\hbar}{2}\boldsymbol{\sigma}_{ab}\psi_{b}$  which results directly from
the definitions of $\mathbf{h}$  and $\psi$, suggests the choice
$\mathbf{S}_{ab}=\frac{\hbar}{2}\boldsymbol{\sigma}_{ab}$; the same result for the ``spin operator''
$\mathbf{S}$ is obtained with the help of the angular momentum commutation relations in QT.
The quantum-like formulation~\eqref{eq:WRZ18THO43Q} of the evolution equations leads
of course to the same result for $g$
as ~\eqref{eq:LTRRU9W2EW},~\eqref{eq:TTOIUOZ14UZ},~\eqref{eq:DTG4EWIU3Z}: The
Zeemann term in~\eqref{eq:WRZ18THO43Q} may be written in the form
$\boldsymbol{\mu_{s}^{op}}\mathbf{B}$, and the relation between $\mathbf{S}$ and the operator of the
magnetic moment $\boldsymbol{\mu_{s}^{op}}$ is given by $\boldsymbol{\mu_{s}^{op}}=
-\frac{g}{2}\frac{e}{mc}\mathbf{S}$ with $g = 2$. This value of the gyromagnetic factor has
been experimentally confirmed to a good approximation (there are corrections to this value
that are not of interest here).

The correct gyromagnetic factor $g = 2$ was obtained by Dirac in 1928, in the course of
the derivation of the relativistic wave equation that bears his name~\cite{dirac:quantum_spin}.
It has long been assumed that the deviation from $g = 1$ is to be regarded as a relativistic (as
well as quantum) effect. About four decades later Levy-Leblond showed, however, that the same
correct value of $g$ may also be derived from Schr\"odinger's equation if a \emph{linearization} with
regard to the differential operator $\partial_{k}$ is performed~\cite{levy-leblond:nonrelativistic}.
One of the assumptions made by Dirac is that the basic quantum equations for a particle
should be linear in all first order derivatives. This assumption may be partly motivated by
the relativistic space-time structure considered by Dirac. It's physical meaning is, however,
not invariably linked to this space-time structure. Lévy-Leblond made the same assumption
as Dirac for the non-relativistic space-time and obtained a differential equation
for a four-component spinor, just as in Dirac's theory. This four-component spinor is composed
of two two-component spinors $\hat{\psi}$ and $\hat{\chi}$ which obey the equations
\begin{gather}
 \big[\frac{\hbar}{\imath} \frac{\partial}{\partial t}+V(q)\big] \hat{\psi}  
 -\sigma_{k}\frac{\hbar}{\imath} \frac{\partial}{\partial q_{k}}\hat{\chi}=0 
\mbox{,}
\label{eq:KUHEWQ89FB}\\
\sigma_{l} \frac{\hbar}{\imath} \frac{\partial}{\partial q_{l}}\hat{\psi}
+2m\hat{\chi}=0
\mbox{.}
\label{eq:FSM21KHFEL}
\end{gather}
The two-spinor $\hat{\chi}$, defined as a linear combination of derivatives of $\hat{\psi}$
with respect to $q_{l}$, plays obviously the role of an auxiliary variable. If $\hat{\chi}$ is
inserted in~\eqref{eq:KUHEWQ89FB} one obtains the field free Pauli-Schr\"odinger equation
if the derivatives of $\hat{\psi}$ with respect to $q_{k}$ commute with each other. If, on the
other hand, a gauge field is turned on with the help of the miminal coupling rule, one obtains
Pauli's equation with the Zeemann term including the correct factor $g=2$.
Lévy-Leblond's derivation shows that spin is not a relativistic phenomenon. This fact is still
not widely known. You can still find statements like: ``Quantum spin arises from the
combination of  special relativity with quantum mechanics''~\cite{lambiase_papini:interaction}. 

The  value $g=2$, which is ``anomalous'' in the sense of a deviation from the result of the
particle image, cannot therefore be explained as a relativistic (quantum mechanical) effect.
This convenient explanation breaks down. Instead, the question now arises why the linearity
in  $\partial_{k}$ should be responsible for this deviation. The answer may be found in an
unjustly forgotten work~\cite{eberlein:spin_model} by Eberlein from 1962. In this work, which
provides the basis for our above considerations, the correct value $g=2$ was found for the
first time in the framework of non-relativistic QT.

As discussed in more detail in section~\ref{sec:two-repr-eucl}, Eberlein introduced a matrix
representation of Euclidean space that is suitable for describing the behavior of spinors. The
deeper reason for this lies in the topology of the rotation group. The result of these topological
considerations agrees with the result of the ``linearization'' (with regard to the differential operator),
which was carried out by Levy-Leblond following Dirac’s method. A comparison of these two
processes shows that this is no coincidence: The conditions for a meaningful ``linearization'' lead
to the same commutation relations as the conditions for a consistent spinor representation of
$\mathbb{R}^{3}$.  This means that the unclear requirement of “linearity” may be replaced by the
obvious requirement that the topological structure of the rotation group has to be correctly taken
into account. But this means that actually no assumption at all is required in order to be able to
derive the correct $g$-factor (it is tempting to associate the factor $2$  with the ratio $4\pi/2\pi$).

This insight automatically leads to the next question. If the deviation of the gyromagnetic factor
from $1$ is a purely topological effect why should we then need QT to derive it. The answer given
by the present theory is that we actually do \emph{not} need QT to derive it; according to the
present derivation, it may be classified as a semiclassical (or semiquantal) effect. The fact that
the Zeeman term is not a quantum effect may also be realized from the very absence of $\hbar$ in
the constant of proportionality between $\boldsymbol{\mu_{s}^{op}}$ and $\mathbf{S}$ (the $\hbar$ that
appears in the definition of $\mathbf{S}$ is irrelevant, as discussed in section~\ref{sec:helic-hopf-invar}).
It should also be mentioned in advance that the transition to QT carried out below does not lead to
any modification of the Zeemann term. The latter is a consequence of the minimal coupling rule if
the momentum field is displayed correctly - that is, taking into account its vortical degrees of freedom. 

Let us now ask for possible interpretations of the “magnetic moment of the electron”. In the particle
picture it was suggested by Uhlenbeck and Goudsmit~\cite{uhlenbeck:spinning}
that an ``intrinsic'' angular momentum of the electron, called \emph{spin}, is responsible for the
observed effects. The magnitude of the spin vector is assumed to be a constant equal to
$\frac{\hbar}{2}$. As already mentioned, the form of the relevant coupling term is compatible
with this assumption, but not the magnitude of the prefactor ($g = 2$ instead of $g = 1$). 
The second, much more serious shortcoming is the universally accepted fact that such a
classical intrinsic spin cannot exist because it is in conflict with fundamental physical
principles~\cite{kronig:spinning}. While the theories of Dirac~\cite{dirac:quantum_spin},
Eberlein~\cite{eberlein:spin_model}, and Levy-Leblond~\cite{levy-leblond:nonrelativistic}
provide an explanation for the value $g = 2$, the fundamental second difficulty remains. 
In a relativistic world the construction of a classical model for a spinning electron is just
as impossible as in the non-relativistic case.Thus, while the existence of quantum spin is
experimentally extremely well confirmed, a classical counterpart of this property - if
interpreted as a property of individual electrons - does not exist~\cite{ohanian:spin}. 
Nevertheless, the intuitive ideas of Uhlenbeck and Goudsmit dominate our thinking about
spin even today. The reason is that there is no better explanation, at least in the framework
of the individuality interpretation (particle picture) of QT. This represents a painful gap in
our understanding of nature, especially in view of the fundamental importance of spin
for the stability of matter. 

This gap in our understanding disappears if one renounces the claim to be able to describe
the behavior of individual particles with the help of QT. In the ensemble interpretation on which the
present work is based, it is possible to explain the origins of quantum spin in a simple way:
The momentum field, which describes a collective of particles and assigns a momentum to
every point of configuration space, must have three independent components, as space is
three dimensional. This simple fact leads to the appearance of two additional fields describing
an \emph{internal degree of freedom of the ensemble} (so one can dispense with the strange
idea that a point particle has internal degrees of freedom). After linearization one obtains the
complete basic equation of non-relativistic QT for a single particle, derived by Eberlein and
Levy-Leblond.

It is of course also possible to introduce additional spin degrees of freedom in phase space.
The problem here, however, is that you then make assumptions that do not correspond to reality.
In contrast, the existence of spin in the ensemble theory follows automatically from the basic
assumption that the dynamic variables of quantum theory only depend on the space-time coordinates
as independent variables.

The two difficulties of the individuality interpretation mentioned above do not arise in the ensemble
interpretation. The difficulty, or better impossibility, to understand spin as an intrinsic angular
momentum of a single particle does not exist in the ensemble theory, since the spin in this interpretation
is a collective property that results from a vortical component in the momentum distribution of the
probabilistic ensemble of \emph{all} particles. Of course, there is then no difficulty in accepting 
a deviation of the gyromagnetic ratio from the single particle value.

\section[Transition to quantum theory by linearization]{Transition to
  quantum theory by linearization}
\label{sec:tran-quant-theo-lin}
Let us briefly recap at this point. Our first step was the projection from phase space to
configuration space,  transforming Eqs.~\eqref{eq:SU2MDVI9ER}-\eqref{eq:DRAE48ZU9EB} to
Eqs.~\eqref{eq:UMW29ISJG},~\eqref{eq:NJZIM2BSEQU},~\eqref{eq:HSE27WG6FDC},~\eqref{eq:NN8M43UQU}.
With this first, most fundamental step, the globally valid theory PM was converted to the only locally
valid theory QA. The next two steps were the introduction of independent variables (potentials) and the transition
to canonical Clebsch potentials. Of course, these two steps, that led to 
Eqs.~\eqref{eq:LBRGR9WEEW}-\eqref{eq:THABIOZ42UZ} do not change anything in the fundamentally
inacceptable nature (only local validity) of the above theory. This also applies to the next step that
we have carried out, namely the rewriting of the basic equations in the quantum-like
form~\eqref{eq:WRO76TDE49Q}.

The form~\eqref{eq:WRO76TDE49Q} of the QA, in which the quantum variable $\psi$ is already
used, is still a theory in which particle trajectories exist - albeit subject to “local validity” as
discussed in III.  Using this form of the basic equations of QA, the influence of an external electromagnetic field
could be taken into account  in a simple way, with the help of the minimal coupling rule.
This form also allows a particularly simple transition to QT. This transition must be performed in
such a way that the global validity, which has been lost during the projection, emerges again from
the locally valid QA. This requires either linearizing Eq.~\eqref{eq:WRO76TDE49Q}  or randomizing
Eqs.~\eqref{eq:LBRGR9WEEW}-\eqref{eq:THABIOZ42UZ}, as discussed
in detail in III. It will be shown that both kinds of ``quantization'' can be carried out quite analogously
to the spinless case. This final transition to QT is a process that is completely independent from the
presence or absense of an electromagnetic field. We therefore omit the electrodynamic terms
in~\eqref{eq:WRZ18THO43Q} for the sake of clarity. 

The transition to QT by linearization takes place in a simple manner by omitting the non-linear term on
the right-hand side of Eq.~\eqref{eq:WRO76TDE49Q}. In the resulting quantum theoretical
evolution equation
\begin{equation}
\label{eq:BRA39TR56Q}
\left[\frac{\hbar}{\imath}\partial_{t}-\frac{\hbar^{2}}{2m}\partial_{k}\partial_{k}+V \right]\psi_{a}= 0
\mbox{,}
\end{equation}
the role of $S,\,\vartheta,\,\varphi$ as functions defining the momentum field is destroyed and
the particle equations of motion $\dot{q}_{k}=v_{k}(q,t)$ become meaningless. All theorems
(such as those of Helmholtz type) that are based on deterministic laws for particle motion
become invalid. This conclusion is of course the same as in the irrotational case studied in III.
Probabilistic theories of this kind, in which probabilistic statements about particles can be made,
while no statements can be made about the orbits of the particles, were referred to as Type 3
theories in an earlier work of the present author~\cite{klein:statistical}. Thus, the restoration of
the global validity of our theory leads to a radical change in its physical meaning. Just as radical
is the change in the mathematical description that is now made possible by linearity. The new
probabilistic structure, the role of eigenvalues and non-commuting observables, Born's rule,
and other characteristics of QT were derived in I and II. 

The equation of motion~\eqref{eq:BRA39TR56Q} is a doubling of the Schr\"odinger equation and
at first glance appears to be equivalent to it. But this is not the case; the new discrete degree of
freedom has important physical consequences even in the absence of an external electromagnetic field.
The latter may be introduced in the same way as in section~\ref{sec:minim-coupl-mathbbe}
and leads to Eq.~\eqref{eq:WRZ18THO43Q} with vanishing right hand side.

\subsection[Alternative description of the linearization	]{Alternative description
  of the linearization}
\label{sec:altern-descr-line}
Let us ask which modifications Eqs.~\eqref{eq:LBRGR9WEEW}-\eqref{eq:THABIOZ42UZ} will
undergo as a consequence of the transition to QT. The answer provides us with a reformulation of
Schr\"odinger's equation in terms of the variables $S,\,\varphi,\,\vartheta,\,\rho$ which will be useful
later. We assume that the new equations take the form
\begin{gather}
\frac{\partial S}{\partial t}(q,t)+ 
\frac{\hbar}{2}\cos\vartheta(q,t) \frac{\partial \varphi}{\partial t}(q,t)+
H^{0}\left(q,\frac{\partial S}{\partial q} + \frac{\hbar}{2}\cos\vartheta \frac{\partial \varphi}{\partial q}\right)=  L_{S}
\mbox{,}
  \label{eq:LB6DT7SFEJ}\\
  \left[\frac{\partial}{\partial t}+V_{k}^{0}\left(q,\frac{\partial S}{\partial q} + \frac{\hbar}{2}\cos\vartheta \frac{\partial \varphi}{\partial q}\right)\frac{\partial}{\partial q_{k}} \right] \varphi(q,t)=L_{\varphi}
\mbox{,}
\label{eq:EB4NZ3RHZQ}\\
\left[\frac{\partial}{\partial t}+ V_{k}^{0}\left(q,\frac{\partial S}{\partial q} + \frac{\hbar}{2}\cos\vartheta \frac{\partial \varphi}{\partial q}\right)\frac{\partial}{\partial q_{k}} \right] \frac{\hbar}{2}\cos\vartheta(q,t)=L_{\vartheta}
\mbox{,}
\label{eq:TKRAGH5OPQ}\\
\frac{\partial \rho}{\partial t}(q,t)+
\frac{\partial}{\partial q_{k}}\rho(q,t)V_{k}^{0}
\left(q,\frac{\partial S}{\partial q} + \frac{\hbar}{2}\cos\vartheta \frac{\partial \varphi}{\partial q}\right)=L_{\rho}
\label{eq:TJ8RHOBA4Z}
\mbox{,}
\end{gather}
with new terms $L_{A}$, where $A = S,\,\varphi,\,\vartheta,\,\rho$, replacing the zeros on the right-hand
sides of Eqs.~\eqref{eq:LBRGR9WEEW}-\eqref{eq:THABIOZ42UZ}.

We will here determine the quantum terms $L_{A}$ in a purely formal way, as a consequence
of the linearization, postponing questions concerning deeper physical meaning.
We denote the left-hand sides of~\eqref{eq:LBRGR9WEEW}-\eqref{eq:THABIOZ42UZ} by $T_{A}$.
The four basic equations $T_{A}=0$ of QA are equivalent to the nonlinear equations
\begin{equation}
  \label{eq:GRKZ95WS2S}
\mathbb{L}\psi+\mathbb{M}=0
  \mbox{,}
  \end{equation}
  where $\mathbb{L}$  and $\mathbb{M}$ are given by
  $\mathbb{L}=\frac{\hbar}{\imath}\partial_{t}-\frac{\hbar^{2}}{2m}\partial_{k}\partial_{k}+V$ and
  $\mathbb{M}_{a}=\frac{\hbar^{2}}{2m}
  \left[f_{k(a)}f_{k(a)}+(\partial_{k}f_{k(a)}) \right]\psi_{a}$.
On the other hand, the four basic equations $T_{A}=L_{A}$ of QT are equivalent to the linear
equations
\begin{equation}
  \label{eq:HSCZ87W3S}
\mathbb{L}\psi=0
  \mbox{.}
  \end{equation}
We multiply the spinor~\eqref{eq:GRKZ95WS2S} from the left by $\psi^{+}$ and $\phi^{+}$,
where the spinor $\phi$, which is orthogonal to $\psi$, is defined by
$ \phi=\sqrt{\rho}\,\mathrm{e}^{\frac{\imath}{\hbar}S}\,\left(u_{2}^{*},-u_{1}^{*} \right)^{T}$. 
Taking advantage of the fact that all equations are evolution equations, containing only a single
first-order derivative with respect to time, we can identify the real and imaginary parts of the
resulting expressions and obtain the relations
\begin{gather}
  \label{eq:GREUZ3P17O}
  \rho T_{S}+\imath\frac{\hbar}{2} T_{\rho}=\psi^{+}\mathbb{M}+\psi^{+}\mathbb{L}\psi\\
    \label{eq:G4IUD3PKH2}
    -\frac{\hbar}{2} \rho T_{\vartheta}+\imath\frac{\hbar}{2}\rho\sin\vartheta T_{\varphi}=
    \phi^{+}\mathbb{M}+\phi^{+}\mathbb{L} \psi
\mbox{.}
  \end{gather}
We proceed in the same way with the spinor~\eqref{eq:HSCZ87W3S} and obtain the relations
\begin{gather}
  \label{eq:AOU3IZ317O}
  \rho \left(T_{S}-L_{S}\right)+\imath\frac{\hbar}{2} \left( T_{\rho}-L_{\rho} \right)= \psi^{+}\mathbb{L}\psi\\
  \label{eq:AOU2D3PLH2}
  -\frac{\hbar}{2} \rho \left( T_{\vartheta}-L_{\vartheta} \right)
  +\imath\frac{\hbar}{2}\rho\sin\vartheta \left( T_{\varphi}- L_{\varphi} \right)=\phi^{+}\mathbb{L} \psi
\mbox{.}
  \end{gather}
Comparison of~\eqref{eq:GREUZ3P17O},~\eqref{eq:G4IUD3PKH2}
and~\eqref{eq:AOU3IZ317O},~\eqref{eq:AOU2D3PLH2} shows that the terms $L_{A}$ obey the equations
$\rho L_{S}+\imath\frac{\hbar}{2} L_{\rho}= \psi^{+}\mathbb{M}$ and
$ -\frac{\hbar}{2} \rho L_{\vartheta} +\imath\frac{\hbar}{2} \rho \sin\vartheta L_{\varphi}=\phi^{+}\mathbb{M}$ .
It is of course reasonable that the additional quantum terms may be determined from the
nonlinear term $\mathbb{M}$  which makes the difference between QA and QT. A longer calculation
leads to the result
\begin{gather}
  L_{S}=\frac{\hbar^{2}}{8m}\Big\{
  4 \rho^{-\frac{1}{2}} \left( \partial_{k}\partial_{k} \rho^{\frac{1}{2}}\right)-
  \sin^{2}\vartheta\left( \partial_{k}\varphi \right) \left( \partial_{k}\varphi \right)-
  \left( \partial_{k}\vartheta \right)     \left( \partial_{k}\vartheta \right)
  \Big\}
\mbox{,}
  \label{eq:LBDRU3HSEJ}\\
L_{\varphi}=\frac{\hbar}{2m}\bigg\{
\cos\vartheta \left( \partial_{k}\varphi \right) \left( \partial_{k}\varphi \right) -
\frac{1}{\sin\vartheta} \left( \partial_{k}\partial_{k}\vartheta \right)-
\frac{1}{\rho\sin\vartheta} \left( \partial_{k}\rho \right) \left( \partial_{k}\vartheta \right) 
  \bigg\}
\mbox{,}
\label{eq:EB4DRUH87EQ}\\
L_{\vartheta}=\frac{\hbar}{2m}\Big\{
2 \rho^{-\frac{1}{2}} \left( \partial_{k}\rho^{\frac{1}{2}}\right) \sin\vartheta \left( \partial_{k}\varphi \right)+
2\cos\vartheta\left( \partial_{k}\varphi \right) \left( \partial_{k}\vartheta \right)+
\sin\vartheta \left( \partial_{k}\partial_{k}\varphi \right)
\Big\}
\mbox{,}
\label{eq:TKRADRUHO2Q}\\
L_{\rho}=0
\label{eq:TJ7DRUHKR2Z}
\mbox{.}
\end{gather}
Equations~\eqref{eq:LB6DT7SFEJ}-~\eqref{eq:TJ8RHOBA4Z}, with the terms $L_{A}$ given
by~\eqref{eq:LBDRU3HSEJ}-~\eqref{eq:TJ7DRUHKR2Z} are equivalent to the
spinorial Schr\"odinger Eq.~\eqref{eq:BRA39TR56Q}. Apart from the continuity equation
\eqref{eq:TJ8RHOBA4Z}, which remains unchanged, the new quantum terms lead
to a coupling of the variables $S,\,\rho,\,\varphi,\,\vartheta$ which makes the concept
of individual particle trajectories meaningless. The situation is basically the same
as in the case of irrotational momentum fields treated in III; in this case only the first term of $L_{S}$
survives, which is sometimes (misleadingly) referred to as “quantum potential”. 
Our next task, performed in section~\ref{sec:tran-quant-theo-rand}, is to understand the
terms $L_{A}$ with the help of statistical concepts.

The form~\eqref{eq:LB6DT7SFEJ}-~\eqref{eq:TJ8RHOBA4Z} of the Schr\"odinger equation, with
the terms $L_{A}$ given by~\eqref{eq:LBDRU3HSEJ}-~\eqref{eq:TJ7DRUHKR2Z}, was derived many
years ago in pioneering work by Bohm and
co-workers~\cite{bohm_schiller_tiomno:causal},~\cite{bohm_schiller:causal} and
Takabayasi~\cite{takabayasi:vortex}. In earlier work by Takabayasi~\cite{takabayasi:vector}
and also in works by Bialynicki-Birula~\cite{bialynicki-birula:hydrodynamic} hydrodynamic
variables (velocity field or momentum field) in addition to spin variables were used as dynamic
variables instead of the three quantum mechanical potentials used in the present theory. The
basic equations of this “hydrodynamic formulation” of QT~\cite{bialynicki-birula:hydrodynamic}
may be derived from the above equations by means of a tedious differentiation.
The derivation of the above equations presented here can be seen as a continuation and completion
of the theories of Bohm and Takabayasi. It has now been possible to derive the same equations
starting from the well-understood classical theory PM, and give also plausible reasons for the
individual steps leading from PM to QT.

\section[Transition to quantum theory by randomization]{Transition to
  quantum theory by randomization}
\label{sec:tran-quant-theo-rand}
The zeroing of the nonlinear term in Eq.~\eqref{eq:WRO76TDE49Q}, that creates QT is, in physical
terms, a randomization. However, it is an unusual kind of randomization. With the standard concept of
randomization, as used in classical probabilistic physics, only the initial conditions are random  while the
particle movement itself is ruled by deterministic laws. This standard type of probabilistic theory was referred to as
type 2 theory in a previous work of the present author~\cite{klein:statistical}. In contrast, the above transition
from~\eqref{eq:WRO76TDE49Q} to~\eqref{eq:BRA39TR56Q} makes the equations of motion
themselves ``random” (which means nonexistent), while statistical predictions about the behavior of
the particles are still possible. This (quantum) type of probabilistic theory was referred to as type 3
theory~\cite{klein:statistical}. The question arises as to whether we can better understand this transition
from type 2 to type 3. Are there statistical assumptions that are equivalent to the linearization process,
explaining, in an alternative way, the transition from~\eqref{eq:WRO76TDE49Q}
[or~\eqref{eq:LBRGR9WEEW}-\eqref{eq:THABIOZ42UZ}] to~\eqref{eq:BRA39TR56Q} ?
In III it was shown that this question can be answered in the affirmative in the case of irrotational momentum
fields. Here we show that the same holds true in the present completed theory. The following construction
is an expanded and improved version of an earlier theory of the present author~\cite{klein:nonrelativistic}. 

\subsection[Definition of Ehrenfest-like relations]{Definition of Ehrenfest-like relations}
\label{sec:defin-ehrenf-like}
There should be as close a relationship as possible between the statistical equations of
type 3 that we want to construct, and the equations of classical mechanics. We assume therefore
the validity of Ehrenfest-like relations of the form 
\begin{gather}
\mathrm{d}_{t}\bar{q}_{k}(t)=\frac{1}{m}\bar{p}_{k}(t)
\mbox{,}
  \label{eq:HEMNU83S2J}\\
\mathrm{d}_{t}\bar{p}_{k}(t)=\overline{F_{k}(q,t)}
\mbox{,}
\label{eq:E18TRATUQ}\\
\mathrm{d}_{t}\bar{s}_{k}(t)=\overline{T_{k}(q,t)}
\mbox{,}
\label{eq:TKHOEGH42Q}
\mbox{,}
\end{gather}
where $F_{k}(q,t)$ is the external force and the “macroscopic variables” $\bar{q},\,\bar{p},\,\bar{s}$
are defined  as average values
\begin{gather}
  \bar{q}_{k}(t)=\int\mathrm{d}q\rho(q,t)q_{k}
\mbox{,}
  \label{eq:HKE3RT75J}\\
  \bar{p}_{k}(t)=\int\mathrm{d}q\rho(q,t)M_{k}(q,t)
\mbox{,}
\label{eq:E1EM4SATUQ}\\
\bar{s}_{k}(t)= \int\mathrm{d}q\rho(q,t)s_{k}(q,t)
\mbox{,}
\label{eq:TKRUTA82Q}
\mbox{,}
\end{gather}
of the “microscopic” field variables $q,\,M,\,s$. The
``statistical conditions''~\eqref{eq:HEMNU83S2J}-\eqref{eq:TKHOEGH42Q}
ensure that the mean values of $M_{k}(q,t)$ and $s_{k}(q,t)$ obey
particle-like relations.  We switch on an external electromagnetic
field by taking into account the terms containing $\Phi, \mathbf{A}$ and $\mathbf{B}$ [see
Eqs.~\eqref{eq:LTRRU9W2EW},~\eqref{eq:BG2FR3TW8Q} and~\eqref{eq:DTG4EWIU3Z}]
which are induced by the process of minimal coupling. We do this because in the course of the following
derivation of Schr\"odinger's equation, it will be possible to establish a connection between
the ``potentials'' $\Phi, \mathbf{A}$ and $\mathbf{B}$ on the one hand and associated
external forces $\mathbf{F}$ on the other hand. It is instructive to write the energy
and momentum components $M_{0}$ and $M_{k}$ in the form
\begin{equation}
  \label{eq:BPAL28JZR3W}
M_{0}=\partial_{t}S^{'}+R_{0},\;\;\;\;\;\;\;\;M_{k}=\partial_{k}S^{'}+R_{k}
  \mbox{,}
\end{equation}
  where
\begin{gather}
  \partial_{t}S^{'}=\partial_{t}S+e\Phi,\;\;\;\;\;\;\;\;\;\partial_{k}S^{'}=\partial_{k}S - \frac{e}{c}A_{k},
  \label{eq:RTUI3R55WJ}\\
  R_{0}=\frac{\hbar}{2}\cos\vartheta\partial_{t}\varphi,\;\;\;\;\;\;\;\;\
  R_{k}=\frac{\hbar}{2}\cos\vartheta\partial_{k}\varphi
  \label{eq:EJ9K4E2ROQ}
  \mbox{.}
\end{gather}
Both the multivalued phase $S^{'}$  (in contrast to $S$ the first derivatives of $S^{'}$ with respect
to $q_{k}$ and $t$ are not interchangeable) and the terms $R_{k}$ contain rotational components
of $M_{k}$. The terms that come from $S^{'}$ describe the external field, and the terms that come
from $R_{k}$ describe the internal part of the rotational component of the momentum field. The
fundamental law of conservation of probability
\begin{equation}
  \label{eq:LOC8OPO2RT}
\partial_{t}\rho+\frac{1}{m}\partial_{k}\rho M_{k}=0
  \mbox{,}
  \end{equation}
with the probability current $v_{k}$ defined by $mv_{k}=M_{k}$, completes the set of our basic quations.
Let us note that the forces $\mathbf{F}$ and $\mathbf{T}$ on the right-hand sides
of~\eqref{eq:E18TRATUQ} and~\eqref{eq:TKHOEGH42Q} have a completely different character.
The form of $\mathbf{T}$ is given by the minimal coupling mechanism
[see Eq.~\eqref{eq:DTG4EWIU3Z}]. In contrast, the form of $\mathbf{F}$ need not
be specified, as will be shown in the next section. 

\subsection[Implementing the statistical conditions]{Implementing the statistical conditions}
\label{sec:impl-stat-cond}
The first statistical condition~\eqref{eq:HEMNU83S2J} is automatically satisfied due to the continuity
equation and the definition of $\bar{q}_{k}(t)$  and $\bar{p}_{k}(t)$ . Using the continuity equation
again and performing some rearrangements, the second statistical condition~\eqref{eq:E18TRATUQ}
takes the form
\begin{align}
  -\int\mathrm{d}q (\partial_{k}\rho)  \left[M_{0} +\frac{1}{2m}M_{i}M_{i} \right]\;\;&  \notag\\
  + \int\mathrm{d}q \rho  \left[ \frac{1}{m}M_{i}\Omega_{ik} + \left(\partial_{t}M_{k}-\partial_{k}M_{0} \right)\right]&
  =\int\mathrm{d}q \rho F_{k}
  \label{eq:AJ9TRATE89Q}
  \mbox{,}
\end{align}
where $\Omega_{ik}$ is the vorticity tensor defined by Eq.~\eqref{eq:VO2OIT9TWE}. 
We have not specified the external forces that appear in our statistical theory. The reason is that the
form of these forces may be \emph{derived} as a consequence of the special structure of our theory. The
Ehrenfest-like relations~\eqref{eq:HEMNU83S2J}-\eqref{eq:TKHOEGH42Q} are integral equations.
Our task is to derive one or more differential equations from these integral equations. The forces
$\mathbf{F}$ on the right-hand side of~\eqref{eq:E18TRATUQ} must be designed to \emph{allow} this. That
means there is a ``statistical constraint'' on the forces in the present theory.

The concrete form of this restriction can already be seen in Eq.~\eqref{eq:AJ9TRATE89Q}: The
permitted forces - that are compatible with the minimal coupling rule - must already appear on
the left-hand side of~\eqref{eq:AJ9TRATE89Q}, namely in the form of statistical mean values. Only
such forces can be real, because they may cancel with the same forces on the right-hand side, thus
disappearing from the integral equation and making he derivation of one or more differential
equations possible.

We now use the decomposition of $M_{0}$ and $M_{k}$ according to Eq.~\eqref{eq:BPAL28JZR3W}.
The square bracket in the second term of Eq.~\eqref{eq:AJ9TRATE89Q} takes the form:
\begin{align}
&\;\;\;\;\;\;\;\;\;\;\;\;\left[ \frac{1}{m}M_{i}\Omega_{ik} + \left(\partial_{t}M_{k}-\partial_{k}M_{0} \right)\right]=\notag\\
&\;\;\;\;\;\;\;\;  \frac{1}{m}(\partial_{i}S^{'}+R_{i})\left[\partial_{i},\partial_{k} \right]S^{'}+
  \left[\partial_{t},\partial_{k} \right]S^{'}\notag\\
& +\frac{1}{m}(\partial_{i}S^{'}+R_{i})\left( \partial_{i}R_{k}- \partial_{k}R_{i}\right)                                            +\partial_{t}R_{k}- \partial_{k}R_{0}
  \label{eq:BG4TRASUZ7Q}
  \mbox{,}
\end{align}
where $\left[\partial_{i},\partial_{k} \right]S^{'}=
\frac{e}{c}\left(\partial_{k}A_{i}-\partial_{i}A_{k} \right)$ and
$\left[\partial_{t},\partial_{k} \right]S^{'}=
-\frac{e}{c}\left(\partial_{t}A_{k}+c\partial_{k}\Phi \right)$. Because of the well-known relations 
$B_{k}=\epsilon_{kij}\partial_{i}A_{j}$ and $E_{k}=-\frac{1}{c}\partial_{t}A_{k}-\partial_{k}\Phi$ 
the first term on the right hand side of~\eqref{eq:BG4TRASUZ7Q} is given by the Lorentz force
\begin{equation}
  \label{eq:H27MWED9DL}
F_{k}^{(L)}=eE_{k}+\frac{e}{c}v_{i}\left( \partial_{k}A_{i}-\partial_{i}A_{k}  \right)
\mbox{.}
\end{equation}
So we write $\mathbf{F}=\mathbf{F}^{(L)} + \mathbf{F}^{(2)}$ and skip the Lorentz force
from both sides of Eq.~\eqref{eq:AJ9TRATE89Q}. The second statistical condition now
takes the form
\begin{align}
  -\int\mathrm{d}q (\partial_{k}\rho)  \left[M_{0} +\frac{1}{2m}M_{i}M_{i} \right]\;\;\;\;\;\;\;\;\;\;\;\;&  \notag\\
  + \int\mathrm{d}q \rho  \left[ +\frac{1}{m}(\partial_{i}S^{'}+R_{i})\left( \partial_{i}R_{k}- \partial_{k}R_{i}\right)+\partial_{t}R_{k}- \partial_{k}R_{0}\right]&
  =\int\mathrm{d}q \rho F_{k}^{(2)}
  \label{eq:AJROUIT271Q}
  \mbox{,}
\end{align}
We now express the vortical components $R_{0},\,R_{k}$ according to~\eqref{eq:EJ9K4E2ROQ}
in terms of the Clebsch potentials and obtain  after a few rearangements the representation
\begin{equation}
  \label{eq:BHHW58THU8Q}
  -\int\mathrm{d}q (\partial_{k}\rho)  \left[M_{0} +\frac{1}{2m}M_{i}M_{i} \right]
-\frac{\hbar}{2}\int\mathrm{d}q \rho \sin\vartheta
\Big\{ (\partial_{k}\varphi)\mathrm{D}_{t}\vartheta  -(\partial_{k}\vartheta)\mathrm{D}_{t}\varphi  \Big\}
  =\int\mathrm{d}q \rho F_{k}^{(2)}
  \mbox{,}
  \end{equation}
for the second statistical condition. At this point the third statistical condition must be taken
into account.

If we identify the minimal coupling force $\mathbf{T}$ with the right hand side
of~\eqref{eq:DTG4EWIU3Z} and use the continuity equation~\eqref{eq:LOC8OPO2RT},
the integral equation~\eqref{eq:TKHOEGH42Q} takes the form
\begin{equation}
  \label{eq:D81RSTBO5D}
\int\mathrm{d}q \rho \left[\mathrm{D}_{t}s_{k}+\frac{e}{mc}\epsilon_{kij}B_{i}s_{j} \right]=0
\mbox{.}
\end{equation}
The trivial solution of this equation (vanishing of the square bracket) agrees with the
quasi-classical field equation~\eqref{eq:DTG4EWIU3Z}. The simplest nontrivial differential
equation to be associated with the integral equation~\eqref{eq:D81RSTBO5D} has the form
\begin{equation}
  \label{eq:FGR7TBU8D}
\mathrm{D}_{t}s_{k}+\frac{e}{mc}\epsilon_{kij}B_{i}s_{j} = \frac{\hbar}{2}G_{k}
\mbox{,}
\end{equation}
where the $G_{k}$  are three functions with vanishing average value. We split off the
factor $\frac{\hbar}{2}$  so that the functions $G_{k}$  do not depend on the parameter
defining the length of the spin vector; the quantities $G_{k}$ should only describe
the influence of the randomization. If we insert $s_{k}$ [see Eq.~\eqref{eq:Z8WEH3DFT9R}]
in~\eqref{eq:FGR7TBU8D} we obtain the differential equations for the two independent
field variables  $\vartheta$  and  $\varphi$,
\begin{gather}
  \mathrm{D}_{t}\vartheta=\frac{e}{mc}
  \left(B_{1}\cos\varphi-B_{2}\sin\varphi \right)-\frac{G_{3}}{\sin\vartheta}
\mbox{,}
\label{eq:HJUNU71Q2J}\\
  \mathrm{D}_{t}\varphi=\frac{e}{mc}\frac{1}{\sin\vartheta}
  \left(-B_{1}\cos\vartheta\sin\varphi-B_{2}\cos\vartheta\cos\varphi +B_{3}\sin\vartheta\right)
+\frac{1}{\sin\vartheta}\left(G_{1}\cos\varphi-G_{2}\sin\varphi \right)
\label{eq:E1PPIRA2U5Q}
\mbox{.}
\end{gather}
The form of these differential equations is still undetermined since the functions $G_{k}$ are not
known. What we do know is that the $G_{k}$ have to obey the conditions
\begin{equation}
  \label{eq:DDII3UZ97EW}
  \int\mathrm{d}q \rho G_{k}=0,\;\;\;\;\;\;\;\;\;G_{k}s_{k}=0
  \mbox{.}
  \end{equation}
 The first of these says that the average value of $G_{k}$  has to vanish, the second
 is a solvability condition that takes into account the fact that $\mathbf{s}$  is a vector
 of constant length. With the help of~\eqref{eq:HJUNU71Q2J} and~\eqref{eq:E1PPIRA2U5Q}
 we can now eliminate the total derivatives $\mathrm{D}_{t}\vartheta$  and $\mathrm{D}_{t}\varphi$
 from the second statistical condition~\eqref{eq:BHHW58THU8Q}. Performing some
 rearrangements and a partial integration Eq.~\eqref{eq:BHHW58THU8Q} takes the form
\begin{align}
  -\int\mathrm{d}q (\partial_{k}\rho)
  \left[M_{0} +\frac{1}{2m}M_{i}M_{i} + \mu_{i}B_{i}\right]
-\int\mathrm{d}q \rho\mu_{i}\partial_{k}B_{i}\;\;\;\;\;\;\;&  \notag\\
  +\frac{\hbar}{2}\int\mathrm{d}q \rho
  \Big\{ G_{1}\cos\varphi \partial_{k}\vartheta-
G_{2}\sin\varphi \partial_{k}\vartheta + G_{3}\partial_{k}\varphi \Big\}
  =\int\mathrm{d}q \rho F_{k}^{(2)}&
  \label{eq:GETW72THA8Q}
  \mbox{,}
  \end{align}
We see that the vortical part of the momentum field leads to the potential energy
term $\mu_{i}B_{i}$, the force $\mu_{i}\partial_{k}B_{i}$, and to the integral depending on
the $G_{i}$. Except for the prefactor, the newly derived force agrees with the
electrodynamic force exerted by an inhomogeneous magnetic field on a magnetic dipole.
As is well known, it plays a central role in the interpretation of the Stern-Gerlach experiment.
We eliminate the new force by setting
\begin{equation}
  \label{eq:DFK3EBKF82V}
F_{k}^{(2)}= -\mu_{i}\partial_{k}B_{i}+ F_{k}^{(3)}
  \mbox{.}
  \end{equation}
The remaining force $F_{k}^{(3)}$ can only have the form $-\partial_{k}V$. This leads to the
usual mechanical potential $V$  in the field equation. Then, the second statistical condition
takes the form
\begin{align}
  -\int\mathrm{d}q (\partial_{k}\rho)
  \left[M_{0} +\frac{1}{2m}M_{i}M_{i} + \mu_{i}B_{i}+V \right]\;\;\;\;\;\;\;\;\;&  \notag\\
  +\frac{\hbar}{2}\int\mathrm{d}q \rho
  \Big\{ G_{1}\cos\varphi \partial_{k}\vartheta-
G_{2}\sin\varphi \partial_{k}\vartheta + G_{3}\partial_{k}\varphi \Big\}=0&
  \label{eq:GUZW82THB8Q}
  \mbox{,}
  \end{align}
The second integral in Eq.~\eqref{eq:GUZW82THB8Q} must be a contribution to the field
equations. This implies a relationship of the form
\begin{equation}
  \label{eq:ARO2TFGU7G}
\frac{\hbar}{2}\int\mathrm{d}q \rho
\Big\{ G_{1}\cos\varphi \partial_{k}\vartheta-G_{2}\sin\varphi \partial_{k}\vartheta +
G_{3}\partial_{k}\varphi \Big\}=\int\mathrm{d}q (\partial_{k}\rho)L_{0}^{'}
  \mbox{,}
  \end{equation}
whereby $L_{0}^{'}$ is an unknown function. Then the second statistical condition takes the form
\begin{equation}
  \label{eq:AHI8TGTRU2G}
 -\int\mathrm{d}q (\partial_{k}\rho)
  \left[M_{0} +\frac{1}{2m}M_{i}M_{i} + \mu_{i}B_{i}+V - L_{0}^{'}\right]=0
  \mbox{.}
  \end{equation}
The simplest nontrivial differential equation whose solutions solve this integral equation is given by
\begin{equation}
  \label{eq:KEW9RTJZW1G}
  M_{0} +\frac{1}{2m}M_{i}M_{i} + \mu_{i}B_{i}+V = L_{0}
  \mbox{,}
  \end{equation}
  where the function $L_{0}$  obeys the conditions
  \begin{equation}
    \label{eq:DB28SGE5UTG}
L_{0}=L_{0}^{'}+\Delta L_{0},\;\;\;\;\;\;\;\;\int\mathrm{d}q (\partial_{k}\rho)\Delta L_{0}=0
    \mbox{.}
    \end{equation}
We have thus transformed the second statistical condition, originally formulated as an
integral equation, into a partial differential equation. The latter contains, however, $4$ 
unknown functions $G_{i},\,L_{0}$ which are still to be determined.

\subsection[Determination of the quantum terms $G_{i},\,L_{0}$]{Determination of the
  quantum terms $G_{i},\,L_{0}$}
\label{sec:determ-quant-terms}
The first fundamental assumption that we made when constructing our type 3 theory was the validity
of the Ehrenfest-like relations~\eqref{eq:HEMNU83S2J}-\eqref{eq:TKHOEGH42Q}. This assumption led
us to the field equations~\eqref{eq:LOC8OPO2RT},~\eqref{eq:KEW9RTJZW1G},~\eqref{eq:HJUNU71Q2J},
\eqref{eq:E1PPIRA2U5Q} for our dynamic variables $\rho,\,S,\,\vartheta,\,\varphi$. These new equations
differ from the field equations of our original type 2 theory by terms $G_{i},\,L_{0}$, which are unknown
functions of our dynamic variables and their derivatives. The conditions listed above are not
sufficient to determine the $G_{i},\,L_{0}$. We need a second assumption, presumably of a
statistical nature. A most fundamental statistical principle says that all states that are unknown
must occur with the same probability. In  statistical mechanics (a type 2 theory) this principle
is implemented through the requirement for maximum entropy. In the problem at hand, we
are faced with the task of determining certain terms in a differential equation in accordance with this
principle. In this case, the same general principle leads to the requirement for minimal Fisher
information~\cite{frieden:fisher_basis},\cite{reginatto:pauli},\cite{klein:statistical},\cite{klein:nonrelativistic}.
In the following we will derive the terms implied by the requirement of minimal Fisher information with
the help of a variational problem. Large parts of the related calculations have already been
reported\cite{klein:nonrelativistic}. In this regard, we may be brief here.

We introduce the abbreviation $L$ for the left hand side of Eq.~\eqref{eq:KEW9RTJZW1G} so that
this equation takes the form $L-L_{0}=0$. We assume that the function $L_{0}$ we are looking for
depends only on the variables $\rho$, $\vartheta$, $\varphi$, and their first and second derivatives
with respect to $q_{k}$~\cite{klein:nonrelativistic}. Our fundamental second statistical assumption
is that the spatial and temporal average value of $L-L_{0}$ is extremal with respect to the
variation of $S,\,\rho,\,\vartheta,\,\varphi$,
\begin{equation}
\delta \int \mathrm{d} t \int \mathrm{d}q   \rho
\left( L - L_{0} \right) = 0 \label{eq:JJ438TRT9RQ} 
\mbox{.}
\end{equation}
Furthermore, the four field
equations~\eqref{eq:LOC8OPO2RT},~\eqref{eq:KEW9RTJZW1G},~\eqref{eq:HJUNU71Q2J},~\eqref{eq:E1PPIRA2U5Q} must also be fulfilled; we write this condition symbolically for the sake of brevity in the form
\begin{equation}
  E_a=0,\;\;\;a=\rho,\,S,\,\vartheta,\,\varphi
\label{eq:HQ3TR4JWQ}
\mbox{.}
\end{equation}
The two conditions~\eqref{eq:JJ438TRT9RQ} and~\eqref{eq:HQ3TR4JWQ} lead to differential
equations for $G_{i},\,L_{0}$. As shown in detail in~\cite{klein:nonrelativistic}, the physically relevant
solution of these equations is given by
\begin{gather}
L_0 = \frac{\hbar^{2}}{2m} \bigg[
\frac{1}{\sqrt{\rho}}
\frac{\partial}{\partial q_{k}} \frac{\partial}{\partial q_{k}}
\sqrt{\rho}
-\frac{1}{4}\sum_{k} \bigg\{ \sin^{2} \vartheta
\left( \frac{\partial \varphi}{\partial q_{k}} \right)^{2}
+ \left( \frac{\partial \vartheta}{\partial q_{k}} \right)^{2}
\bigg\}
\bigg]
\mbox{,}
 \label{eq:DIKRALU27HO} \\
\hbar G_1 = \frac{\hbar^{2}}{2m} \frac{1}{\rho}
\frac{\partial}{\partial q_{k}} \rho
\bigg( \frac{1}{2} \sin 2\vartheta \sin \varphi
\frac{\partial \varphi}{\partial q_{k}}
- \cos \varphi \frac{\partial
\vartheta}{\partial q_{k}}
\bigg)
\mbox{,}
\label{eq:HER2NALZ7HO} \\
\hbar G_2 = \frac{\hbar^{2}}{2m} \frac{1}{\rho}
\frac{\partial}{\partial q_{k}} \rho
\bigg( \frac{1}{2} \sin 2\vartheta \cos \varphi
\frac{\partial \varphi}{\partial q_{k}}
+ \sin \varphi \frac{\partial
\vartheta}{\partial q_{k}}
\bigg)
\mbox{,}
 \label{eq:D8SHL3OI2H}\\
\hbar G_3 = - \frac{\hbar^{2}}{2m} \frac{1}{\rho}
\frac{\partial}{\partial q_{k}}
\big( \rho \sin^{2} \vartheta \frac{\partial \varphi}{\partial q_{k}} \big)
\label{eq:D9TELGU12W}
\mbox{.}
\end{gather}
A new adjustable parameter appears on the right-hand-sides of these expressions which
has been identified with $\hbar^{2}/2m$.  Let us recall here that two different adjustable
parameters appeared in the course of our developments both of which were identified with
$\hbar$. The first  $\hbar$ was associated with the length of the spin vector $\mathbf{s}$.
The second  $\hbar$ is associated with the quantum-mechanical principle of maximal disorder.
The physical meaning of these two adjustable parameters is different, but they must both be
identified with Planck's constant in order to enable the transition to QT.

As may be easily checked, the
conditions~\eqref{eq:DDII3UZ97EW},~\eqref{eq:ARO2TFGU7G},~\eqref{eq:DB28SGE5UTG}
for the $L_{0},\,G_{i}$ are all fulfilled. In particular, $L_{0}=L_0^{\prime}+\Delta L_0$, where
\begin{equation}
  \label{eq:AE9K3EUZ3F}
L_0^{\prime} =
-\frac{\hbar^{2}}{8m} \bigg[ \sum_{k} \bigg\{ \sin^{2} \vartheta
\left( \frac{\partial \varphi}{\partial q_{k}} \right)^{2}
+\left( \frac{\partial \vartheta}{\partial q_{k}} \right)^{2}
\bigg\}
\bigg],
\;\;\;\;\;
\Delta L_0 =\frac{\hbar^{2}}{2m}\frac{1}{\sqrt{\rho}}
\frac{\partial}{\partial q_{k}} \frac{\partial}{\partial q_{k}}\sqrt{\rho}
\mbox{,}
\end{equation}
and $L_0^{\prime}$ fulfills~(\ref{eq:ARO2TFGU7G}). One can also show that the mean value
of $L_{0}$ agrees with the Fisher functional~\cite{reginatto:pauli},~\cite{klein:nonrelativistic}.
If the above solution for $L_{0}$  is inserted, then the variational principle~\eqref{eq:JJ438TRT9RQ}
leads to the correct field equations~\eqref{eq:HQ3TR4JWQ}. We did not use the principle of
variation here in the usual way, as a mathematical tool to derive field equations from a given
Lagrangian function, but we use it instead to construct the Lagrangian itself.

Finally, we should compare the above field theory, which is based on statistical postulates, with
the earlier one generated by linearization. The easily verifiable relations
\begin{gather}
L_{S}=L_{0},
 \label{eq:EJ83BWQD2A}\\ 
L_{\vartheta}=-\frac{1}{\sin\vartheta}G_{3},
  \label{eq:RDRHZ316WJ}\\
L_{\varphi}=\frac{\cos\varphi}{\sin\vartheta}G_{1}- \frac{\sin\varphi}{\sin\vartheta}G_{2}
  \label{eq:EJ9HNZRO7A}
  \mbox{.}
\end{gather}
show that both theories are identical. We have thus shown that one may obtain the Pauli equation
\begin{equation}
  \label{eq:HAB87THJK3Q}
  \left[\frac{\hbar}{\imath}\partial_{t}-e\Phi-\frac{\hbar^{2}}{2m}
 \sum_{k=1}^{3} \, \left( \partial_{k}-\imath\frac{e}{\hbar c}A_{k}\right)^{2}
    +\mu_{B}\sigma^{k}B_{k}+V \right] \psi=0
  \mbox{.}
  \end{equation}
either through the “formal” process of linearization or through the implementation of some
plausible statistical postulates. The first way is a discontinuous process that destroys
the possibility of describing particle motion and thus has the physical meaning of a randomization.
With the second way, we have replaced this discontinuous process with a continuous one.
This enabled us to understand the detailed nature of this randomization. The second way is
longer and ``less elegant'' than the first, but allows a more precise insight into the transition from
QA to QT.

\section[Discussion]{Discussion}
\label{sec:discussion}
The second version of the HLLK reported here starts from PM, generates the theory QA by projection onto
configuration space, and then realizes  the transition to QT by a linearization or randomization. The
theory QA, appearing only in this second version of the HLLK, represents the  transition area,
the ``borderland'', between classical physics and QT~\cite{berry:classical}. This theory is unphysical
in that the particle trajectories that occur are only locally valid and are not realized in nature (although
they may be a good approximation in certain situations). The importance of this theory is that it represents
a clearly defined and justifiable intermediate step in the construction of the QT.

Let us first summarize what has been achieved so far by looking at the list of essential properties of
QT already given in II:
\begin{enumerate}
\item Schr\"odinger's equation as fundamental dynamical law - and eigenvalues as observable numbers.
\item The nonstandard probabilistic structure of QT - in particular non-commuting observables. 
\item Born's rule - the law which tells us how to extract probabilistic predictions from the theory.
\item The minimal-coupling rule - the way interactions are formulated in QT.
\item The existence of spin - a particularly mysterious phenomenon believed to belong to QT exclusively
\item The anomalous value of the magnetic moment of the electron - a spin related phenomenon
\item The spin-statistics connection - a spin related multi-particle phenomenon
\end{enumerate}
Points 1, 2 and 3 were derived in papers I and II. Without going into detail, we mention that the
generalization of Born’s rule to degenerate states may be obtained in a straightforward way (the classical
counterpart of degenerate states are non-connected level sets). In contrast to I and II, only a single
observable, namely the Hamilton function $H(q,p)$, is studied in III and the present work.
Our research in the ``borderland'' created a new derivation of the Schr\"odinger equation, which
allows for a deeper understanding of the relationship between QT and classical physics.
We found, as most important result of the present work, that spin is not a purely
quantum mechanical phenomenon and that the value $1/2$, which is experimentally observed
for the spin of all massive structureless particles, is a consequence of the three-dimensionality
of space. An associated result is the value $g=2$ for the gyromagnetic ratio the electron. We
have thus essentially understood points $5$ and $6$. In the course of our second derivation of
the Schr\"odinger equation (in which statistical assumptions were used) we were also able to
understand why in QT the influence of external force fields must be described with the help
of potentials. Of course, we do not claim to have explained all the details related to points $1-6$.
The last point $7$, the spin-statistics connection for massive particles, remains open at the
moment, but we have little doubt that this point can also be explained in the framework of the HLLK.
In the remainder of this section, we continue to discuss some relevant points in more detail.

\subsection[The meaning of spin]{The meaning of spin}
\label{sec:meaning-spin}
\epigraph{Pls tell me why is spin of an electron +1/2 or -1/2 and not something like +1 or -1?}
{\textit{- Tridib Banerjee (age 16) Kolkata, India}~\cite{tridib:spin}}
In this work it was possible to identify the origin of spin. This property is not, as has
long been assumed, to be regarded as a purely quantum mechanical phenomenon.
Rather, its origin may be found in the quasi-classical (or quasi-quantum) theory QA,
which represents the borderland between classical physics and QT. As this theory is not
realized in nature it would not be correct to claim that spin has a classical origin. There is
no such phenomenon in phase space (although it is formally possible to introduce
corresponding degrees of freedom) since no self-rotation of individual particles exists in nature.

The possibility of discovering the origin of spin arises by replacing the particle momenta
$p$ by momentum fields $M (q, t)$, and thus halving the number of degrees of freedom.
A necessary prerequisite for the occurrence of spin is the simple fact that all degrees of
freedom of the momentum fields $M (q, t)$ have to be taken into account, in particular those
which result in a non-vanishing rotation. The origin of the quantum spin is thus a collective
property of the probabilistic ensemble in configuration space, which is associated with the
rotational degrees of freedom. The idea that spin should better be understood as a
non-localized phenomenon has already been expressed several times in the
literature~\cite{ohanian:spin},~\cite{chuu:semiclassical}.

A simple and extremely important property of nature is that spin-$1/2$ particles are fermions,
which means they obey Fermi-Dirac statistics. When asked about this fact, Richard Feynman suggested
that there should be a simple explanation for such a fundamental fact. And that if we cannot find
such a simple explanation, we should admit that we have not understood the phenomenon.
We cannot give an answer to Feynman’s question in the present work. But we can answer an
even more fundamental question, a question so fundamental that it is hardly ever asked by
professionals. Namely the question, why all massive structureless particles in nature have spin $1/2$.
Based on the present reconstruction of QT for the case of a single particle, the reason
is the three-dimensionality of space. As a consequence of this three-dimensionality we have to
introduce \emph{three} functionally independent potentials in order to represent the momentum
field in a correct way. Together with the probability density, we have four independent real functions
that is a $2$-component spinor, in other words a spin $1/2$ particle. This answer is hard to
beat in terms of simplicity.

According to the current state of knowledge in the theory of elementary particles, (almost) all
structureless massive particles are actually spin $1/2$ particles. One would justifiably object
that reality has a relativistic space-time structure that is not correctly described by our
non-relativistic theory. But it is also a fact that our conclusion is essentially based only on
the number of spatial dimensions. It is therefore not unreasonable to assume that our conclusion
remains valid even when moving to a relativistic theory.

\subsection[The limit $\hbar \to 0$ of the Pauli-Schrödinger equation]{The limit $\hbar \to 0$
  of the Pauli-Schrödinger equation}
\label{sec:limit-hbar-to}
In this section we discuss the question of which theory the Pauli-Schrödinger theory “reduces” to
in the limit $\hbar \rightarrow 0$.  There are two types of reduction which Rosaler called
formal and empirical reduction~\cite{rosaler:formal}; a similar distinction was made by
Berry~\cite{berry:classical}. The important difference between these two concepts was
discussed in detail in III. Here we are only examining the concept of formal reduction, that is,
we are examining the behavior of the basic equations of our theory in the limit of small  $\hbar$.

The above question was decided for a long time on the basis of philosophical postulates.
Also, no distinction was made between formal and empirical reduction. The first
mathematically precise investigation of the question of formal reduction was carried out
in~\cite{klein:what} and II and led to the conclusion that QT (without spin) cannot be reduced
to either CM or PM. This conclusion meant that - quite contrary to the prevailing opinion - QT
cannot (formally) be reduced to classical physics, but, on the contrary, it must be possible to
derive it from classical physics. This conclusion thus provided the basis for the reconstruction
of QT carried out in I-III and here. The general conclusion that QT  reduces for small  $\hbar$
neither to CM nor to PM but to a (quasi-classical) probabilistic theory in configuration space,
does not depend on whether we take spin into account or not. The presence of spin leads
however to some peculiarities, which we want to summarize in the following.

The version of the Pauli-Schrödinger equation that is best suited for examining the limit of 
small $\hbar$ is given by
Eqs.~\eqref{eq:KEW9RTJZW1G},~\eqref{eq:FGR7TBU8D},~\eqref{eq:LOC8OPO2RT}
for the potentials $S,\,\vartheta,\,\varphi$ and the probability density $\rho$. To make the
dependence on $\hbar$ visible, we write Eqs.~\eqref{eq:KEW9RTJZW1G},~\eqref{eq:FGR7TBU8D}
in the form
\begin{gather}
\partial_{t} S+e\Phi+ 
\frac{\hbar}{2}\cos\vartheta\, \partial_{t} \varphi+\frac{1}{2m}\sum_{k}
\left( \partial_{k}S-\frac{e}{c}A_{k}+ \frac{\hbar}{2}\cos\vartheta\, \partial_{k} \varphi \right)^{2}
-\frac{\hbar}{2}\frac{e}{mc} s_{i}B_{i}=  L_{0}
\mbox{,}
\label{eq:LR3LAMEE5Z}\\
\left[\partial_{t}+\frac{1}{m} \left( \partial_{i}S-\frac{e}{c}A_{i}+ \frac{\hbar}{2}\cos\vartheta\ (\partial_{i}\varphi) \right)    \partial_{i}\right] h_{k}+\frac{e}{mc}\epsilon_{kij}B_{i}h_{j} = G_{k}
\label{eq:E7AL6UMAH4}
\mbox{,}
\end{gather}
where $L_{0}$  and $G_{k}$ are of the order $\hbar^{2}$  and $\hbar$, respectively and the mechanical
potential has been ommitted. We do not write down the continuity equation, which behaves in an
obvious way. If we set $\hbar = 0$ in these equations, we obtain the equations
\begin{gather}
\partial_{t} S+e\Phi+\frac{1}{2m}\sum_{k}\left( \partial_{k}S-\frac{e}{c}A_{k}\right)^{2}=  0
\mbox{,}
\label{eq:LREGMHI73Z}\\
\left[\partial_{t}+\frac{1}{m} \left( \partial_{i}S-\frac{e}{c}A_{i} \right) \partial_{i}\right] h_{k}
+\frac{e}{mc}\epsilon_{kij}B_{i}h_{j} = 0
\label{eq:E7KRE8G68AH4}
\mbox{,}
\end{gather}
which, together with the continuity equation, represent the classical limit of the Pauli-Schrödinger
equation. This is a classical (deterministic) field theory defined by the Hamilton-Jacobi
equation~\eqref{eq:LREGMHI73Z} for a spinless charged particle in an electromagnetic field
and the equation of motion~\eqref{eq:E7KRE8G68AH4} of the variable $h_{k}$  which is associated
with the vortical component of the momentum field. The change of $h_{k}$ with time is determined
by the solution $S$  of~\eqref{eq:LREGMHI73Z}, while conversely~\eqref{eq:LREGMHI73Z} does
not depend on $h_{k}$ .

The “survival” of spin variables in the case $\hbar = 0$ is no surprise in our theory, since we
have identified the vortical components of the momentum field as origin of quantum spin.
In all works in which Eqs.~\eqref{eq:LR3LAMEE5Z},~\eqref{eq:E7AL6UMAH4} were derived
so far, the starting point was the quantum mechanical Pauli-Schrödinger
equation~\eqref{eq:HAB87THJK3Q}, which was then rewritten, using a representation
like~\eqref{eq:DH92RLK2VP}
(see \cite{takabayasi:vortex},~\cite{bohm_schiller_tiomno:causal},~\cite{yahalom:fluid_spin}).
The limiting case $\hbar = 0$ was rarely dealt with in these theories, see
however~\cite{yahalom:fluid_spin}. One reason for this might be that this limiting
case is not compatible with the prevailing interpretation of spin as a purely
quantum mechanical phenomenon. Due to this interpretation, all spin variables (or
the corresponding terms in a Lagrangian function) should disappear from the theory in the limit
$\hbar = 0$. The fact that this is not the case led Yahalom to the conclusion that the Pauli
theory ``has no standard classical limit''~\cite{yahalom:fluid_spin}. In fact, one could have
concluded from this fact that spin cannot be a purely quantum mechanical phenomenon.

A second difficulty is as follows. If one wants to eliminate only terms of order $\hbar^{2}$ in
Eqs.~~\eqref{eq:LR3LAMEE5Z},~\eqref{eq:E7AL6UMAH4} and keep all terms of order $\hbar$,
then one finds that this is not possible, since it destroys the structure of the kinetic energy.
A meaningful theory, namely the Eqs.~\eqref{eq:LTRRU9W2EW}-\eqref{eq:BG2FR3TW8Q} which
were the starting point for our last step to QT, can only be obtained if one eliminates only the
terms $L_{0}\,\hbar G_{k}$  and keeps terms of any order in $\hbar$ that come from the spin
amplitude. The present derivation from classical physics provides us automatically with a
physically meaningful distinction between these two kinds of $\hbar^{2}$ terms.

\subsection[The role of potentials]{The role of potentials}
\label{sec:role-potentials}
Why does the minimal coupling rule apply in QT? The point where it appears in
our formalism is the projection from phase space to configuration space.
This projection introduces an energy field $M_{0}$ and three components of the
momentum field $M_{k}$  [we may write these in the simplest case as derivatives
with respect to $t$ and $q_{k}$ of an action variable $s(q,t)$, see
Eq.~\eqref{eq:NN8M43UQU}]. An external field can then no longer be taken into account
by introducing a force field, as in Newton's equations. Such an external influence can
actually only be taken into account by modifying the fields $M_{0}$ and $M_{k}$. The
minimal coupling rule represents the simplest possible modification of these fields, namely
the \emph{addition} of externally specified fields (potentials) $e\Phi$  and $-\frac{e}{c}A_{k}$
to $M_{0}$ and $M_{k}$. Konopinski has already pointed out that potentials provide field
energy and momentum for exchange with charged matter~\cite{konopinski:vector_potential}.
In the context of the present derivation, there seems to be almost no alternative to this
interpretation.

In the present framework of the ensemble interpretation of QT, it is clear
that a local argument based on potentials is questionable. Rather, the non-gauge-invariant
quantities $e\Phi$  and $-\frac{e}{c}A_{k}$  should be interpreted as describing the influence
of an external electromagnetic field on a statistical ensemble as a whole. This fact
requires a fundamental reconsideration of the Aharonov-Bohm effect, which will not be
undertaken here. 

\subsection[Concerning the interpretation]{Concerning the interpretation}
\label{sec:concerning-inter}
Large parts of the scientific community are still dominated by the idea that the
“single-particle Schrödinger equation” describes the behavior of a single particle. This
idea is incompatible with the fact that one can only derive \emph{statistical predictions}
about the behavior of a single particle from the single-particle Schrödinger equation.
All attempts to remove this logical contradiction with the help of complicated constructions
have been unsuccessful and always will be. It can only be removed by abandoning
certain philosophical principles, such as the belief that any “complete” description of nature
must be deterministic. Then one can accept that the one-particle Schrödinger equation
describes only a statistical ensemble of individual particles. This statistical interpretation
forms the basis for the theory described in this series of works, which allows for an almost
complete reconstruction of QT. This reconstruction includes not only the formal aspects
but also all essential questions of the interpretation of the formalism. This high degree
of agreement is a strong argument in favor of the ensemble interpretation.
\begin{acknowledgements}
Open access funding provided by Johannes Kepler University Linz.
\end{acknowledgements}

\bibliographystyle{spphys}       % APS-like style for physics
\bibliography{uftbig}
\end{document}